\documentclass[fleqn,a4paper,12pt]{article}
\usepackage{amssymb}
\usepackage{makeidx}
\usepackage{amsmath}
\usepackage{graphicx}
\usepackage{lscape}
\usepackage{a4}
\usepackage{epsfig}

\renewcommand{\bar}{\overline} \newcommand{\newc}{\newcommand}
\newc{\beq}{\begin{equation}} \newc{\eeq}{\end{equation}}
\newc{\bea}{\begin{array}} \newc{\eea}{\end{array}}
\newc{\ri}{{\mathrm i}}
\newc{\bW}{{\mathbf W}}
\newc{\bR}{{\mathbf R}}
\newc{\bN}{{\mathbf N}}
\newc{\Psibar}{\overline\Psi}
\newc{\w}{{\bf w}}
\newc{\E}{{\mathbf{E}}}
\newc{\bp}{{\bf p}}
\newc{\ta}{\tilde a}
\newc{\bV}{{\bf V}}
\newc{\bfV}{{\bf V}}
\newc{\bfG}{{\bf G}}
\newc{\bx}{{\bf x}}
\newc{\bu}{{\bf u}}
\newc{\bP}{{\bf P}}
\newc{\bJ}{{\bf J}}
\newc{\bK}{{\bf K}}
\newc{\pd}{{\partial}}
\newc{\ti}{{\times}}
\newc{\bA}{{\bf A}}
\newc{\bE}{{\bf E}}
\newc{\bfn}{{\bf\nabla}}
\newc{\ho}{\hookrightarrow}
\newc{\ra}{\rightarrow}
\newc{\bv}{{\bf v}}
\newc{\bb}{{\bf b}}
\newc{\bc}{{\bf c}}
\newc{\bd}{{\bf d}}
\newc{\tbb}{\tilde{\bf b}}
\newc{\tbc}{\tilde{\bf c}}
\newc{\tbd}{\tilde{\bf d}}
\newc{\bz}{{\bf 0}}
\newc{\bun}{{\bf 1}}
\newc{\bL}{{\bf L}}
\newc{\bS}{{\bf S}}
\newc{\bB}{{\bf B}}
\newc{\br}{{\bf r}}
\newc{\sig}{{\mathbf\sigma}}
\newc{\eg}{{\it e.g.\ }}
\newc{\bpi}{{\mathbf\pi}}
\newc{\ie}{{\it i.e.\ }}
\newc{\etal}{{\it et al}}

\def\JPA#1#2#3#4{#2 #1 {\em J. Phys. A: Math. Gen.} {\bf #3} #4}

\def\AM#1#2#3#4{#2 #1 {\em Ann. Math.} {\bf #3} #4}

\def\NCB#1#2#3#4{#2 #1 {\em Nuov. Cim.} {\bf #3 B} #4}

\def\PRD#1#2#3#4{#2 #1 {\em Phys. Rev.} {\bf D #3} #4}

\def\IJTP#1#2#3#4{#2 #1 {\em Int. J. Theor. Phys.} {\bf #3} #4}
\def\CMP#1#2#3#4{#2 #1 {\em Comm. Math. Phys.} {\bf #3} #4}

\def\TMP#1#2#3#4{#2 #1 {\em Theor. Math. Phys.} {\bf #3} #4}

\def\TMP#1#2#3#4{#2 #1 {\em Theor. Math. Phys.} {\bf #3} #4}

\def\JPG#1#2#3#4{#2 #1 {\em J. Phys. G: Nucl. Part. Phys.} {\bf
#3} #4}

\catcode`@=11 \long
\def\@caption#1[#2]#3{\par\addcontentsline{\csname
ext@#1\endcsname}{#1} {\protect\numberline{\csname
the#1\endcsname}{\ignorespaces #2}} \begingroup \small
\@parboxrestore \@makecaption{\csname fnum@#1\endcsname}
{\ignorespaces #3}\par \endgroup} \catcode`@=12

\begin{document} \begin{titlepage} \vskip 2cm
\begin{center} {\Large\bf \Large\bf Galilei invariant theories.\\II.
Wave equations for massive fields} \footnote{E-mail:
{\tt niederle@fzu.cz},\ \ {\tt nikitin@imath.kiev.ua} } \vskip 3cm
{\bf  J. Niederle$^a$, \\ and A.G. Nikitin$^b$ } \vskip 5pt {\sl
$^a$Institute of Physics of the Academy of Sciences of the Czech
Republic,\\ Na Slovance 2, 18221 Prague, Czech Republic} \vskip 2pt
{\sl $^b$Institute of Mathematics, National Academy of Sciences of
Ukraine,\\ 3 Tereshchenkivs'ka Street, Kyiv-4, Ukraine, 01601\\}
\end{center}
\vskip .5cm \rm

\begin{abstract} Galilei invariant equations for massive fields  with various spins
 are found and classified. They have been obtained
directly, i.e., by using requirement of Galilei invariance and the
facts on representations of the Galilei group deduced in our
previous paper de Montigny M,  Niederle J and Nikitin A G, J. Phys.
A {\bf 39}, 1-21, 2006 . It is shown that the collection
 of non-equivalent Galilei-invariant wave equations for vector and scalar
 fields is very broad and  describes many physically consistent systems.

\end{abstract}

\end{titlepage}

\setcounter{footnote}{0} \setcounter{page}{1}
\setcounter{section}{0}

\section{Introduction}

It is well known that
the Galilei group $G(1,3)$ and its representations play the same role in
non-relativistic physics as the Poincar\'e group $P(1,3)$ and its
representations do in the relativistic one.  In fact the
Galilei group and its representations form a
group-theoretical basis of classical mechanics and
electrodynamics. They replace the Poincar\'e group
and its representations whenever velocities of bodies are much smaller then that of light in
vacu\-um. On the other hand, the structure of subgroups of the Galilei group
and of its representations are in many respects more complex than those
of the Poincar\'e group and therefore it is  not perhaps
so surprising that the representations of the Poincar\'e group were
described in \cite{Wigner3} almost 15 years earlier than the representations
of the Galilei
group \cite{bargman} in spite of the fact that the relativity principle of
classical physics was
formulated by Galilei in 1632, about three centuries prior that of
relativistic physics by Einstein.

An excellent review of representations of the Galilei
group was written by L\'evy-Leblond \cite{levyleblond}. It appears
that, as distinct from the Poincar\'e group, that the Galilei group has besides ordinary
also  projective representations (see
\cite{bargman} and \cite{Wigner} respectively). However its subgroup -- the
homogeneous Galilei group $HG(1,3)$ which plays in non-relativistic
physics the role of the Lorentz group in the relativistic case, has a
more complex structure so that its finite-dimensional
indecomposable representations are not classifiable in general (for details see
\cite{Marc}). And they are the representations which play a key role
in formulation of physical models satisfying the Galilei relativity
principle!

An important class of indecomposable finite-dimensional
representations of $HG(1,3)$ was found and completely
classified in \cite{Marc}.  It contains all representations of
the homogeneous Galilei group which when restricted to
its rotation subgroup,  decompose to spin 0, 1/2 and 1 representations. Moreover,
a connection of these representations with those of the Lorentz group by means of the
In\"on\"u-Wigner
contractions was cleared up in \cite{Marc} and \cite{NN2} too.

This offers possibilities to construct various quantum
mechanical and field theoretical models for interacting particles
and
fields with spins 0, 1/2 and 1. For instance, the most general Pauli interaction of Galilean spin--1/2 particles with
an external electromagnetic field can be found in
\cite{Marc}.

In the present paper we study the vector and spinor
representations of the homogeneous Galilei group in detail and use them for
construction various wave equations for particles with spin 0, 1/2, 1 and 3/2.

There are well--developed theories of wave equations invariant
w.r.t. the Poincar\'e group which can be taken for a start of
construction of the Galilei invariant equations.  First, we begin with
the Bhabha approach which is a direct extension of the method yielding the
Dirac equation. The corresponding relativistic wave equations can be written as
systems of linear first order partial-differential equations of the form:
\beq
\label{equ}\left(\beta_\mu p^\mu+\beta_4m\right)\Psi({\bf x}, t)=0,
\eeq where $p^0=i\frac\partial{\partial x_0},\
p^a=i\frac\partial{\partial x_a}\  (a=1,2,3),$ and $\beta_\mu\
(\mu=0,1,2,3)$ and $\beta_4$ are square matrices restricted by the condition
of the Poincar\'e invariance. Notice that in the relativistic approach
matrix $\beta_4$ is usually assumed to be proportional to a unit
matrix.

The theory of the Poincar\'e-invariant equations (\ref{equ}) is
clearly explained for instance in the Gel'fand--Minlos--Shapiro book
\cite{gelfand}. Some particular results related to the
Galilei invariant equations (\ref{equ}) can be also found in
\cite{levyleblond}, \cite{Marc}-\cite{ourletter}. Galilean analoques of
Bargman-Wigner equations are presented in the recent paper
\cite{Mar}. However, these equations became incompatible whenever
the minimal interaction with an external e.m. field be introduced
\cite{Mar}.

The other approaches make use of the tensor calculus, and the
associated equations have the form of covariant vectors or tensors
(see, e.g., \cite{corson}). Popular examples of these equations are
the Proca \cite{proca},
 Rarita-Schwinger \cite{rarita} and
Sign--Hagen \cite{sign} equations. Let us mention that all these equations
violate causality or predict incorrect values of the gyromagnetic ratio $g$. The tensor-spinorial equations for particles with
arbitrary half-integer spin which are not violating causality and admit the right value for $g$ have been
 discussed in detail in \cite{niederle}.

In the present paper we use both above mentioned approaches and
derive the Galilei-invariant equations for particles with spins 0,
$\frac12$, 1 and $\frac32$. Moreover we present a complete
description of Galilei-invariant equations (\ref{equ}) for scalar
and vector fields.


\section{Galilei algebra and Galilei--invariant wave equations}

In this section we use a Galilean version of the Bhabha
approach and present a complete list of the corresponding Galilei-invariant wave
equations.

Equation (\ref{equ}) is said to be {\it invariant} w.r.t. the Galilei
transformations \beq\label{11}\bea{l} t\to t'=t+a,\\{\bf x}\to {\bf
x}'= {\bf R}{\bf x} +{\bf v}t+\bf b,\eea\eeq  where $a,{\bf b}$,
${\bf v}$ are real parameters and $\bf R$ is a rotation matrix, if
function $\Psi$ in (\ref{equ}) cotransforms as\beq\label{cova} \Psi({\bf x},
t)\to\Psi'({\bf x'}, t')=e^{\ri f({\bf x}, t)}T\Psi({\bf x}, t),\eeq i.e.,
according to a particular representation of the Galilei group.
Here  $T$ is a matrix depending on transformation
parameters only,  $f({\bf x}, t)=m\left({\bf v} \cdot{\bf
x}+t{v^2}/{2}+c\right)$, $c$ is an arbitrary constant and $\Psi'({\bf x'}, t')$ satisfies the
same equation in prime variables as $\Psi({\bf x}, t)$ in the initial ones.

 The Lie algebra corresponding to representation (\ref{cova}) has the
following generators
\beq\label{cov}\bea{l} P_0=\ri\partial_0,\ \  P_a=-\ri\partial_a,\ M=mI,\\
J_a=-\ri\varepsilon_{abc}x_b\partial_c+S_a,\\
G_a=-\ri x_0\partial_a-mx_a+\eta_a,\end{array}\eeq where $S_a$ and
$\eta_a$ are matrices which satisfy the following commutation
relations: \begin{gather}\label{e3}
[S_a,S_b]=\ri\varepsilon_{abc}S_c,\\\label {e33}
{[}\eta_a,S_b{]}=\ri\varepsilon_{abc}\eta_c,\
{[}\eta_a,\eta_b{]}=0\end{gather} that is, they form a basis of $hg(1,3)$ -- of the
Lie algebra of the homogeneous Galilei group.

Let us note that special classes of the Galilei invariant equations
(\ref{equ})  were described in \cite{nikitin} and  \cite{ourletter}.

Equation (\ref{equ}) is invariant with respect to the Galilei
transformations (\ref{cova}),  if their generators (\ref{cov})
transform  solutions of  (\ref{equ}) into solutions. This
requirement together with the existence of the Galilei invariant
Lagrangian for (\ref{equ}) yields the following conditions on
matrices $\beta_\mu$ ($\mu=0,1,2,3,$) and $\beta_4$ \cite{fuschich}:
\beq\bea{l}\label{cond}
\eta_a^\dag\beta_4-\beta_4\eta_a=-i\beta_a,\\
\eta_a^\dag\beta_b-\beta_b\eta_a=-i\delta_{ab}\beta_0,\\
\eta_a^\dag\beta_0-\beta_0\eta_a=0,\ \  a,b=1,2,3.\eea\eeq

Moreover, $\beta_0$ and $\beta_4$ should to be scalars w.r.t.
rotations, i.e., they have to commute with $S_a$.

Thus the problem of classification of the Galilei invariant
equations (\ref{equ}) is equivalent to find the matrices $S_a, \
\eta_a, \ \beta_0, \beta_a$ and $\beta_4$ satisfying relations
(\ref{e3}),  (\ref{e33}) and (\ref{cond}). Unfortunately, a
subproblem of this problem, i.e., the complete classification of
non-equivalent finite--dimensional representations of algebra
(\ref{e3}), (\ref{e33}) appears to be in general unsolvable (that is a `wild'
algebraic problem). However, for two important particular cases,
i.e., for the purely spinor and vector-scalar representations
the problem of finding of all finite--dimensional indecomposable representations of algebra $hg(1,3)$
was completely solved in
\cite{Marc} .

\subsection{Spinor fields and the corresponding wave equations}

Let $\tilde s$ be the highest value of spin which appears when
representation of algebra $hg(1,3)$ is reduced to a direct sum of
irreducible representations of its subalgebra $so(3)$ generated by $S_a$ which satisfy (\ref{e3}).
Then the
corresponding  representation space of this representation of $hg(1,3)$ is said to be a space of
fields of spin $\tilde s$.

As mentioned in \cite{Marc} there exist only two
non-equivalent indecomposable representations of algebra $hg(1,3)$
defined on the fields of spin 1/2. One of them, $D_{\frac12}(1)$,
when restricted to the subalgebra $so(3)$ remains irreducible while
the other one, $D_\frac12(2)$, decomposes to two irreducible
representations $D(1/2)$ of $so(3)$.  The
corresponding matrices $S_a$ and $\eta_a$ can be written in the
following form: \beq\label{1/2a}
S_a=\frac12\sigma_a \texttt{ and } \eta_a=\bz \texttt{ for } D_\frac12(1)\eeq and
\beq\label{02}
S_a=\frac12\left(\bea{cc}\sigma_a&\bz\\\bz&\sigma_a\eea\right)
 \texttt{ and }
\eta_a=\frac{\ri}2\left(\bea{cc}\bz&\bz\\\sigma_a&\bz\eea\right)  \texttt{ for }D_\frac12(2).\eeq
Here $\sigma_a$ are the Pauli matrices and $\bz$ is the $2\times2$
zero matrix.

Realization (\ref{1/2a}) with conditions
(\ref{cond}) yields  equation (\ref{equ}) trivial, i.e., with zero $\beta$--matrices.

The elements of the carrier space of representation (\ref{02}) will be
called Galilean bi-spinors. It can be found in \cite{Marc} how the
Galilean bi-spinors transform w.r.t.  the finite transformations
from the Galilei group (see also equation (\ref{fin}) below).

Solutions of relations (\ref{cond}) with $S_a, \eta_a$ given by
formulae (\ref{02}) can be written as
\beq\beta_0=\left(\bea{cc}\texttt{I}&\bz\\\bz&\bz\eea\right), \
\beta_a=\left(\bea{cc}\bz&\sigma_a\\\sigma_a&\bz\eea\right),\
a=1,2,3,\ \beta_4=\left(\bea{cc}\kappa \texttt{I}&-\ri\omega
\texttt{I}\\\ri\omega \texttt{I}&2\texttt{I}\eea\right),\label{beta}
\eeq where $\texttt{I}$ and $\bz$ are the 2$\times$2 unit and zero
matrices respectively, and $\omega$ and $\kappa$ are constant multiples.

 Notice that parameter $\kappa$ can
be
 chosen zero since the transformation $\Psi\to
e^{\ri\kappa mt}\Psi$ lives equation (\ref{equ}) invariant. Parameter $\omega$ is inessential too since
it can be annulled by the transformation $\beta_\texttt{m}\to U^\dag
\beta_\texttt{m} U,$ where $\texttt{m}=0,1,2,3,4$ and
\[U=\left(\bea{cc}\texttt{I}&\ri \omega \texttt{I}\\\bz&\texttt{I}\eea\right).\]

Galilean invariance of the equation (\ref{equ}) with $\beta_\texttt{m}$ given in (\ref{beta}) can be
verified by using the explicit transformations
(\ref{11}) and (\ref{cova}) for the space-time variables and wave function
respectively, where
\beq\label{fin}T=\exp(\ri{\mbox{\boldmath$\eta\cdot v$\unboldmath}})
\exp(\frac\ri2{\mbox{\boldmath$\sigma\cdot \theta$\unboldmath}}) =\left(1+
{\mbox{\boldmath$\eta\cdot v$\unboldmath}}\right)\left(\cos\frac{\theta}{2}
+\ri\frac{\mbox{\boldmath$\sigma\cdot \theta$\unboldmath}}{\theta}
{\sin\frac{\theta}{2}}\right
).\eeq
Here ${\mbox{\boldmath$\theta$\unboldmath}}=(\theta_1,\theta_2,\theta_3)$ are
rotation parameters, $\theta=\sqrt{\theta_1^2+\theta_2^2+\theta_3^2}$,
and {\mbox{\boldmath$\eta$\unboldmath}} is a matrix three-vector whose
components  $\eta_1,\eta_2, \eta_3$ are given in (\ref{02}).

Let us note that if
we consider a more general case in which matrices $S_a$ and $\eta_a$ are
represented by a direct sum of an arbitrary finite number of
matrices (\ref{02}) and solve the related equations (\ref{cond}),
then we obtain matrices $\beta_\texttt{m}$ which can be reduced to direct
sums of matrices (\ref{beta}) and zero matrices. In other words,
equation (\ref{equ}) with matrices (\ref{beta}) is the only
non-decoupled system of the first order equations for spin 1/2 field
invariant under the Galilei group.

Equation (\ref{equ}) with matrices (\ref{beta}) and
$\omega=\kappa=0$ coincides with the L\'evy-Leblond equation in
\cite{levyleblond}.

Let us remark that matrices
$\hat\gamma_\texttt{n}=\eta\beta_\texttt{n}|_{\kappa=\omega=0},\
\texttt{n}=0,1,2,3,4 $ with
\beq\label{geta}\eta=\left(\bea{cc}\bz&\texttt{I}\\\texttt{I}&\bz\eea\right)\eeq
satisfy the following relations \beq\label{klifford}
\hat\gamma_\texttt{n}\hat\gamma_\texttt{m}+\hat\gamma_\texttt{m}\hat\gamma_\texttt{n}=2\hat
g_{\texttt{nm}},\eeq where $\hat g_{\texttt{nm}}$ is a symmetric
tensor whose non-zero components are \beq\label{g}\hat g_{04}=\hat
g_{40}=-\hat g_{11}=-\hat g_{22}= -\hat g_{33}=1.\eeq

In the Galilei--invariant approach tensor (\ref{g}) plays the same
role as the metric tensor for the Minkovski space in the relativistic
theory.

The matrices $\hat\gamma_\texttt{m}$ will be used many times later
on. Therefore, for convenience, we present them explicitly:
\beq\label{gamma}\hat\gamma_0=\left(\bea{cc}\bz&\bz
\\\texttt{I}&\bz\eea\right),\
\
\hat\gamma_a=\left(\bea{cc}\bz&-\sigma_a\\\sigma_a&\bz\eea\right),\
\ \ a=1,2,3,\ \
\hat\gamma_4=\left(\bea{cc}\bz&2\texttt{I}\\\bz&\bz\eea\right).\eeq

\subsection{Vector fields and the corresponding wave equations}
\subsubsection{Indecomposable representations for vector fields}

A complete description of indecomposable representations of algebra
$hg(1,3)$ in the spaces of vector and scalar fields is given in
\cite{Marc}. The corresponding matrices $S_a$ and $\eta_a$ have the
following forms:
\beq\label{s}S_a=\left(\bea{cc}\texttt{I}_{n\times n}\otimes
s_a&\cdot
\\
\cdot&\bz_{m\times m} \eea\right),\ \ \eta_a
=\left(\bea{cc}A_{n\times n}\otimes s_a&B_{n\times m}\otimes k^\dag_a\\
C_{m\times n}\otimes k_a &\bz_{m\times m}\eea\right),\eeq where
$\texttt{I}_{n\times n}$ and $\bz_{m\times m}$ are the unit and zero
matrices of dimension ${n\times n}$ and ${m\times m}$ respectively,
$A_{n\times n}$ $B_{n\times m}$ and $C_{m\times n}$ are matrices of
the indicated dimensions whose forms will be specified later on,
$s_a$ are
 matrices of spin one with elements $(s_a)_{bc}=i
\varepsilon_{abc}$, $k_a$ are $1\times3$ matrices of the form
\beq\label{k} k_1=\left (\ri, 0, 0\right),\qquad k_2=\left (0, \ri,
0\right), \qquad k_3=\left (0, 0, \ri\right).\eeq

Matrices (\ref{s}) fulfill relations (\ref{e3}) and (\ref{e33}), iff
matrices $A_{n\times n}, B_{n\times m}$ and $C_{m\times n}$ satisfy
the following relations (we omit the related subindices):
\beq\label{A1} AB=0,\ \ CA=0,\ \ A^2+BC=0.\eeq This system of matrix
equations appears to be completely solvable, i.e. it is possible to
find all non-equivalent indecomposable matrices $A, B$ and $C$ which
satisfy relations (\ref{A1}). Any set of such matrices generates a
representation of algebra $hg(1,3)$ whose basis elements are given
by equations (\ref{s}).

According to \cite{Marc} indecomposable representations
$D(n,m,\lambda)$ of $hg(1,3)$ for scalar and vector fields are
labelled by integers $n, m$ and $\lambda$. They specify dimensions
of submatrices in
(\ref{s}) and the rank of matrix $B$ respectively. As shown in
\cite{Marc}, there exist ten non-equivalent irreducible
representations which correspond to matrices $A_{n\times n}, \
B_{n\times m}$ and $C_{m\times n}$ given in the Table 1.

In addition to the scalar representation whose generators are
presented in formula (\ref{s}) and Item 1 of the table, there exist
nine vector representations corresponding to matrices enumerated in
the Table 1, items 2-10. The related basis elements are the matrices of
dimension $(3n+m)\times(3n+m)$ whose explicit forms are given in
(\ref{s}) and Items 1-10 of  Table 1.

\begin{center}
Table 1. Solutions of equations (\ref{A1})
\end{center}

\begin{tabular}{|l|l|l|}
\hline
$\texttt{No}$&$(n,m,\lambda)$&$ \texttt{Matrices }A,\ B,  C$\\
\hline
1.&(0,1,0)&$A,\ B\ \texttt{and} \ C\ \texttt{do not exist since }m=0$\\
2.&(1,0,0)&$A=0, B\ \texttt{and} \ C\ \texttt{do not exist since }n=0$\\
3.& (1,1,0) &$A=0,\ B=0,\ C=1$\\
4.& (1,1,1) &$A=0,\ B=1,\ C=0$\\
5.& (1,2,1)&$A=0,\ B=(1\ 0),\
C=\left(\bea{c}0\\1\eea\right)$\\
6.&(2,0,0)&$A=\left(\bea{cc}0&0\\1&0\eea\right),\
B\ \texttt{and}\ C\ \texttt{do not exist since }n=0$\\&&\\
7.&(2,1,0)&$A=\left(\bea{cc}0&0\\1&0\eea\right),\
B=\left(\bea{c}0\\0\eea\right), \ C=(1\ 0)$\\&&\\
8.&(2,1,1)&$A=\left(\bea{cc}0&0\\1&0\eea\right),\
B=\left(\bea{c}1\\0\eea\right), \ C=(0\ 0)$\\&&\\
9.& (2,2,1)&$A=\left(\bea{cc}0&0\\1&0\eea\right),\
B=\left(\bea{cc}0&0\\1&0\eea\right),\
C=\left(\bea{cc}0&0\\1&0\eea\right)$
\\&&\\
10.&(3,1,1)&
$A=\left(\bea{ccc}0&0&0\\1&0&0\\0&1&0\end{array}\right),\ \
B=\left(\bea{r}0\\0\\-1\eea\right),\ C=(1\ 0\ 0)$\\\hline
\end{tabular}

\vspace{2mm}

The finite Galilei transformations of vector fields (which can be
obtained by integrating the Lie equations for generators (\ref{s})) and
examples of such fields can be found in paper \cite{Marc}. Here we present
two examples of Galilei vectors which have been in fact  already used in
Section 2.

The matrix five-vector $\hat\gamma=(\hat\gamma_0,\ \hat\gamma_1,\
\hat\gamma_2,\ \hat\gamma_3, \ \hat\gamma_4)$, whose components are
given by equation (\ref{gamma}), form a carrier space of
representation $D(1,2,1)$. Under a Galilei boost its components
transform as $\hat\gamma_\texttt{m}\to T(0,{\bf v})
\hat\gamma_\texttt{m} T^{-1}(0,{\bf v})$, where $T(0,{\bf v})$
 are transformation matrices (\ref{fin}) for ${\mbox{\boldmath$
 \theta$\unboldmath}}\equiv0$. The explicit form of these transformations
 is:
 \beq\label{gtrans}\hat\gamma_0\to \hat\gamma_0,\ \hat
 \gamma_a\to\hat\gamma_a+v_a
 \hat\gamma_0,\ a=1,2,3,\ \hat\gamma_4\to\hat\gamma_4+
 {\bf v}\cdot{\mbox{\boldmath$\hat\gamma$\unboldmath}}
 +\frac{v^2}{2}\hat\gamma_0.\eeq

 This five--vector $\hat\gamma$ is involved in the spinor equation
 (\ref{equ}) with $\beta_\texttt{m}=\eta\hat\gamma_\texttt{m}$. Another
 five-vector used there has the following form
 \beq\label{p}p=(p^0,\  p^1,\ p^2,\ p^3,\ p^4),\eeq
 where
 \beq\label{pcomp}p^0=\ri\frac{\partial}{\partial t},\
 p^a=\ri\frac\partial{\partial x_a},\ \ a=1,2,3,\ \text{and}\  p^4=m.\eeq

 The Galilei transformation law for $p$ is analogous to (\ref{gtrans}),
 namely
 \beq\label{ptrans}p_0\to\tilde p^0=p^0+{\bf p}\cdot {\bf v}+ \frac{v^2}{2}
 p^4,\ \ \
 {\bf p} \to \tilde{\bf p}={\bf p}+{\bf v}p^4,\ \texttt{and } p^4\to\tilde p^4= p^4.\eeq
 Notice that relations (\ref{ptrans}) are in accordance with (\ref{11}) and
 (\ref{cova}), namely, $\tilde p^\texttt{n}=
 \exp(-\ri f(t',{\bf x}'))\\\times p'^{\texttt{n}}\exp(\ri f(t',{\bf x}')),\
  \texttt{n}=0,1,2,3,4$,
  where
 $p'^0=\ri\frac{\partial}{\partial t'}$  and $ p'^a=\ri\frac\partial{\partial
 x'_a}$.
The convolution $\hat\gamma_\texttt{m} p^\texttt{m}$ is a Galilean scalar, and consequently equation (\ref{equ}) with matrices (\ref{cova}) is Galilei--invariant.

\subsubsection{General wave equations for vector and scalar fields}

Let us consider equation (\ref{equ}) and describe all admissible matrices
$\beta_4$ compatible with the invariance conditions (\ref{cond})
. We shall
restrict ourselves to matrices $\eta_a, S_a$ belonging to
representations described in Subsection 2.1 (see Table 1) or to
direct sums of these representations. Then the general form of
matrices $S_a$ and $\eta_a$ is again given by equations (\ref{s})
where, however, matrices $A,B$ and $C$ can be reducible:
\beq\label{n1}A=\left(\bea{ccccc}A_1&&&&\\&A_2&&&\\&&\cdot&&\\&&&\cdot&
\\&&&&\cdot\eea\right),\ B=\left(\bea{ccccc}B_1&&&&\\&B_2&&&\\&&\cdot&&\\
&&&\cdot&\\&&&&\cdot\eea\right),\
C=\left(\bea{ccccc}C_1&&&&\\&C_2&&&
\\&&\cdot&&\\&&&\cdot&\\&&&&\cdot\eea\right)\\\eeq
and are of dimensions $N\times N, M\times N $ and $N\times M$ respectively
with $N$
and $M$ being arbitrary integers. The unit and zero matrices in the associated spin
operator ${\bf S}$ defined by equation (\ref{s}) are of dimension
$N\times N$ and $M\times M$ respectively.

The sets of matrices $(A_1, B_1, C_1), \ (A_2, B_2, C_2), ...$ are
supposed to be indecomposable sets presented in Table 1. Any of
them is labelled by a multiindex $q_i=(n_i,m_i,\lambda_i),\ i=1,2,\cdots$.

Matrices $\beta_4$ and $\beta_0$ should commute with $\bf S$ and therefore
have the following block diagonal form:
\beq\label{n2}\beta_4=\left(\bea{cc}R_{N\times N}&\bz_{N\times M}\\
\bz_{M\times N}&E_{M\times M}\eea\right),\ \beta_0=\left(\bea{cc}
F_{N\times N}&\bz_{N\times M}\\
\bz_{M\times N}&G_{M\times M}\eea\right).\eeq

Let us denote by $|q,s,\nu>$ a vector belonging to a carrier space
of representation $D_q$ of algebra $hg(1,3)$, where
$q=(n,m,\lambda)$ is a multiindex which labels a particular indecomposable
representation as indicated in Table 1, $s$ is a spin quantum number which is equal to 0,1
and index $\nu$ specifies degenerate subspaces with the
same fixed $s$. Then taking into account that matrix
$\beta_4$  commutes with $S_a$  its elements can be expressed as \beq\label{b1}<q,s,\lambda|\beta_4|q',s',\lambda'>=\delta_{s1}
\delta_{s'1}
R_{\lambda\lambda'}(q,q')+\delta_{s0}\delta_{s'0}E_{\lambda\lambda'}(q,q').
\eeq

In order to find matrices $R(q,q')$ and $E(q,q')$ (whose elements are
denoted by
$R_{\lambda\lambda'}$ and $E_{\lambda\lambda'}$ respectively) expression  (\ref{b1}) should be substituted
 into (\ref{cond}) and matrices $\eta$ in form (\ref{s}) together with relations (\ref{A1}) used. As a result we obtain the
following condition \beq\label{b2}(A^\dag)^2R+R(A')^2=A^\dag
RA'-C^\dag EC',\eeq where $A,\ C$ (and $A',\ C'$) are submatrices
used in (\ref{A1}), which correspond to representation $D_q$ (and
$D_{q'}$).

Formulae (\ref{b1}) and (\ref{b2}) express all necessary and
sufficient conditions for matrix $\beta_4$ imposed by the Galilei
invariance conditions (\ref{cond}). Suppose a matrix $\beta_4$
(\ref{b1}) satisfying (\ref{b2}) be known, then the remaining matrices
$\beta_a$ ($a=1,2,3$) and $\beta_0$ can be found by a direct use of
the first and second relations (\ref{cond}). By this way we obtain
\beq\label{b3}\bea{l}<q,s,\lambda|\beta_0|q',s',\lambda'>=
\delta_{s1}\delta_{s'1}
F_{\lambda\lambda'}(q,q')+\delta_{s0}\delta_{s'0}
G_{\lambda\lambda'}(q,q'),\\
<q,s,\lambda|\beta_a|q',s',\lambda'>=i\delta_{s1}\delta_{s'1}
H_{\lambda\lambda'}(q,q')s_a+\delta_{s1}\delta_{s'0}
M_{\lambda\lambda'}(q,q')k_a^{\dag}\\
\ \ \ \ \ \ \ \ \ \ \ \ \ \ \ \ \ \ \ +\delta_{s0}\delta_{s'1}N(q,q')_{\lambda\lambda'}k_a,\eea\eeq where
$s_a$ are matrices of spin one,  $k_a$ are matrices (\ref{k}) and
$F(q,q'), \ G(q,q'), \ H(q,q'),$ $ \ M(q,q'),\ N(q,q')$ are matrices
defined by the following relations \beq\label{b4}\bea{l}H=A^\dag
R-RA',\ \ M=C^\dag E-RB',\ \
N=B^\dag R-EC',\\
F=C^\dag EC' +A^\dag RA',\ \  G=2B^\dag RB'-B^\dag C^\dag E-EC'B'.
\eea\eeq

Thus to derive a Galilei-invariant equation (\ref{equ}) for
vector fields it is sufficient to choose a realization of algebra
$hg(1,3)$ from Table 1 or a direct sum of such realizations, and
find the associated matrix $\beta_4$ (\ref{b1}) whose block matrices
$R$ and $E$ satisfy relations (\ref{b2}). Then
the corresponding matrices $\beta_0$ and $\beta_a$ (\ref{b3}) can be
found from (\ref{b4}).

All non-trivial solutions for matrices $R$ and $E$ are specified in the
Appendix.

\subsubsection{Equations for  scalar and vector fundamental particles}

In the previous section we have found all matrices $\beta_\texttt{m}$ for which
equation
(\ref{equ}) is invariant with respect to vector and scalar
representations of the homogeneous Galilei group. However, the Galilei
invariance itself guaranties neither the consistency
nor the right
number
of independent equations.  Moreover the spin content
of the obtained equations as well as their possibilities to
describe fundamental quantum mechanical systems have not been discussed.

A non-relativistic quantum system is said be {\it fundamental} if the space
of its states forms a carrier space of an irreducible representation of
the Galilei group $G(1,3)$. We shall call such systems
"non-relativistic particles" or simply "particles".

For the group $G(1,3)$ there exist the following three invariant operators
\beq\label{cas1}C_1=M,\ C_2=2MP_0-{\bf P}^2,\ \texttt{and
}C_3=(M{\bf J}- {\bf P}\times {\bf G})^2.\eeq Here ${\bf P, J}$ and
${\bf G}$ are three-vectors whose components are specified by equations
(\ref{cov}). Eigenvalues of these operators are associated with
mass, internal energy and square of mass multiplied by eigenvalue of total spin operator respectively.

Galilei-invariant equation (\ref{equ}) is said to be {\it consistent} and
describes a particle with mass $m$, internal energy $\varepsilon$
and spin $s$ if it has  non-trivial solutions $\psi$ which form a carrier
space for a representation of the Galilei group in which the
following conditions \beq\label{cas2} C_1\psi=m\psi,\ \
C_2\psi=\varepsilon \psi,\ \ C_3\psi=m^2s(s+1)\psi\eeq
are true.

We stress that relations (\ref{cas2}) generate extra conditions for $\beta$--matrices so that
equation (\ref{equ}) with such $\beta_\texttt{m}$ guaranties the validity of equations (\ref{cas2}).

Using definitions (\ref{cov}) we find the following forms
of the invariant operators
\beq\label{cas3}\bea{l}C_1=Im,\ \ C_2=2mp_0-{\bf p}^2,\\
C_3=m^2{\bf S}^2+m({\bf S}\times {\mbox{\boldmath$\eta$\unboldmath}})
\cdot{\bf p}-m({\mbox{\boldmath$\eta$\unboldmath}}\times{\bf S}
)\cdot{\bf p}+{\bf p}^2{\mbox{\boldmath$\eta$\unboldmath}}^2- ({\bf
p}\cdot{\mbox{\boldmath$\eta$\unboldmath}})^2.\eea\eeq

We see that $C_3$ is a rather complicated second-order differential
operator with matrix coefficients. In order to diagonalize this operator, we apply the similarity transformation
\beq\label{cas4}\psi\to \psi'=W\psi,\ C_a\to C_a'=WC_aW^{-1},\ \
a=1,2,3,\eeq with
\beq\label{W0}W=\exp\left(\frac{i}{m}{\mbox{\boldmath$\eta$\unboldmath}}
\cdot{\bf p}\right).\eeq

Since $({\mbox{\boldmath$\eta$\unboldmath}}\cdot{\bf
p})^3=0$ for representations $D(3,1,1)$ and $D(1,2,1)$ and
$({\mbox{\boldmath$\eta$\unboldmath}}\cdot{\bf p})^2=0$ for the
other representations described in Sections 2.1 and 2.2.1, $W$ is the
second- or the first-order differential operator in $\bf x$. Using conditions
(\ref{cond}) we find that \beq\label{cas5}C_1'=C_1,\ C_2'=C_2, \
C_3'=m^2{\bf S}^2\eeq and consequently for consistent equations $\psi'$ satisfies (\ref{cas2}) with the transformed invariant operators (\ref{cas5}), i.e.,
\beq\label{cas6}(2mp_0-{\bf p}^2)\psi'=\varepsilon \psi'\eeq
and
\beq\label{cas61} {\bf S}^2\psi'=s(s+1)\psi'.\eeq

In order to see when conditions (\ref{cas6}) and (\ref{cas61}) are true
we transform  equation (\ref{equ}) by means of $W$. We obtain
\beq\label{cas7} \left(\beta_0C_2 -\beta_42m^2\right)\psi'=0\eeq
since in accordance with (\ref{cond})
\[L'=(W^{-1})^\dag 2m(\beta^\mu p_\mu
-\beta_4 m)W^{-1}=\beta_0(2mp_0-{\bf p}^2) -\beta_42m^2.\]

In order (\ref{cas7}) to be compatible with (\ref{cas6}) the matrix
$M=\beta_0\varepsilon-\beta_42m^2$ (where $\varepsilon $ is an arbitrary parameter) should be non-regular. Moreover,
solutions of equation (\ref{cas7}) must also satisfy condition
(\ref{cas61}) and form a carrier space of irreducible representation
$D(s)$ of the rotation group; consequently equation (\ref{cas7})
must have $(2s+1)$ independent solutions.

As shown in Section 2.2, both matrices $\beta_0$ and $\beta_4$ has the block diagonal form given
by equation (\ref{n2}). Thus equation (\ref{cas7}) is decoupled to two subsystems
\beq\label{cas8}\left(RC_2-2m^2F\right)\varphi_1=0,\eeq and
\beq\label{cas9}\left(EC_2-2m^2G\right)\varphi_2=0\eeq
where the functions $\varphi_1$ and $\varphi_2$
are columns with $3n$ and $m$ components respectively such that
$\psi'=\texttt{column}\left(\varphi_1,\
\varphi_2\right)$ .

Equations (\ref{cas8}) and (\ref{cas9}) describe a particle of spin
$s=1$  (and internal energy $\varepsilon$) provided matrices $R, F,
E$ and $G$ satisfy the following conditions:
\beq\label{cas10}\bea{l}\texttt{Rank}||\varepsilon R-2m^2F||=n-1,\ \
\ \ \texttt{Rank}||\varepsilon E-2m^2G||=m.\eea\eeq

In the case of the equation for particle of spin $s=0$ we have
instead of (\ref{cas10}) the following relations
\beq\label{cas11}\bea{l}\texttt{Rank}||\varepsilon R-2m^2F||=n,\ \ \
\ \texttt{Rank}||\varepsilon E-2m^2G||=m-1.\eea\eeq

Thus to find Galilei invariant equations (\ref{equ}) for particle
with a fixed spin we have to take into account equations discussed in
the previous section as well as conditions (\ref{cas10}) or
(\ref{cas11}) imposed on block components of
matrices $\beta_0$ and $\beta_4$.

The explicit forms of matrices $\beta_4,\ \beta_0 $ and $\beta_a$ for
the indecomposable representations of algebra $hg(1,3)$ are presented in the next subsection.

\subsubsection{Equations invariant with respect to the indecomposable
 representations}

Let us restrict ourselves to the indecomposable representations of
algebra $hg(1,3)$ specified in equation (\ref{s}) and Table 1, and find
the associated matrices $\beta_4,\ \beta_0 $ and $\beta_a$ which appear
in the Galilei invariant equations (\ref{equ}).  Taking into account
that $A'=A,\ B'=B$ and $C'=C$ in (\ref{b2}),  where $A,\ B$ and $C$ are matrices given in
Table 1, we easily find the associated block matrices $R$, $E$ and consequently all
matrices (\ref{b4}). To simplify matrices $\beta_\texttt{m}\ (\texttt{m}=0,1,\cdots, 4)
$ we used equivalence transformations
$\beta_\texttt{m}\to W^\dag\beta_\texttt{m} W,$ where $W$ are invertible matrices
commuting with the Galilei boost generators $\eta_a$.

It appears that non-trivial solutions for
$\beta_4$ (and consequently also for $\beta_0, \beta_a$) exist only for
representations $D(1,1,0)$, $D(2,1,0)$, $D(2,2,1)$ and $D(3,1,1)$. They have
 the following form: \begin{gather}\label{M1}\bea{l}
\texttt{Representation } D(1,1,0):\\\beta_4=\left(\bea{cc}\texttt{I}_{3\times3}&\bz_{3\times1}\\
\bz_{1\times3}&0\eea\right),\
\beta_0=\left(\bea{cc}\bz_{3\times3}&\bz_{3\times1}\\
\bz_{1\times3}&2\eea\right),\
\beta_a=\ri\left(\bea{cc}\bz_{3\times3}&-k_a^\dag\\
k_a&0\eea\right),\eea\\\label{M2} \bea{l}
\texttt{Representation }D(2,1,0):\\
\beta_4=\left(\bea{ccc}\bz_{3\times3}&\bz_{3\times3}&\bz_{3\times1}\\
\bz_{3\times3}&\texttt{I}_{3\times3}&\bz_{3\times1}\\\bz_{1\times3}&
\bz_{1\times3}&1 \eea\right),\ \beta_0=\left(\bea{ccc}2
\texttt{I}_{3\times3}
&\bz_{3\times3}&\bz_{3\times1}\\
\bz_{3\times3}&\bz_{3\times3}&\bz_{3\times1}\\\bz_{1\times3}&
\bz_{1\times3}&0
\eea\right),\ \ \ \ \ \  \\\\
\beta_a=\ri\left(\bea{ccc}\bz_{3\times3}
&s_a&k_a^\dag\\
-s_a&\bz_{3\times3}&\bz_{3\times1}\\-k_a& \bz_{1\times3}&0
\eea\right),\eea\end{gather}
\[ \bea{l}
\texttt{Representation }D(2,2,1):\\
\beta_4=\left(\bea{cccc}\bz_{3\times3}&\bz_{3\times3}
&\bz_{3\times1}&\bz_{3\times1}\\
\bz_{3\times3}&\texttt{I}_{3\times3}&\bz_{3\times1}&\bz_{3\times1}\\
\bz_{1\times3}&
\bz_{1\times3}&0&0\\
\bz_{1\times3}& \bz_{1\times3}&0&1 \eea\right),\
\beta_0=\left(\bea{cccc}2 \texttt{I}_{3\times3}&\bz_{3\times3}
&\bz_{3\times1}&\bz_{3\times1}\\
\bz_{3\times3}&\bz_{3\times3}&\bz_{3\times1}&\bz_{3\times1}\\
\bz_{1\times3}&
\bz_{1\times3}&2&0\\
\bz_{1\times3}& \bz_{1\times3}&0&0\eea\right),\eea\]\beq\label{M3}
\beta_a=\ri\left(\bea{cccc}\bz_{3\times3}&s_a
&\bz_{3\times1}&k_a^\dag\\
-s_a&\bz_{3\times3}&-k_a^\dag&\bz_{3\times1}\\
\bz_{1\times3}&
k_a&0&0\\
-k_a& \bz_{1\times3}&0&0 \eea\right),\eeq
\[ \bea{l}
\texttt{Representation }D(3,1,1):\\
\beta_4=\left(\bea{cccc}\bz_{3\times3}&\bz_{3\times3}
&\nu \texttt{I}_{3\times3}&\bz_{3\times1}\\
\bz_{3\times3}&\nu \texttt{I}_{3\times3}&\texttt{I}_{3\times3}&\bz_{3\times1}\\
\nu \texttt{I}_{3\times3}&
\texttt{I}_{3\times3}&\bz_{3\times3}&\bz_{3\times1}\\
\bz_{1\times3}& \bz_{1\times3}&\bz_{1\times3}&-\nu \eea\right),\
\beta_0=\left(\bea{cccc}\bz_{3\times3}&\texttt{I}_{3\times3}
&\bz_{3\times3}&\bz_{3\times1}\\
\texttt{I}_{3\times3}&\bz_{3\times3}&\bz_{3\times3}&\bz_{3\times1}\\
\bz_{3\times3}&
\bz_{3\times3}&\bz_{3\times3}&\bz_{3\times1}\\
\bz_{1\times3}&
\bz_{1\times3}&\bz_{1\times3}&0\eea\right),\eea\]

\beq\label{M4}\bea{l}
\beta_a=\ri\left(\bea{cccc}\bz_{3\times3}&\bz_{3\times3}
&s_a&\bz_{3\times1}\\
\bz_{3\times3}&\bz_{3\times3}&\bz_{3\times3}&-k_a^\dag\\
-s_a&
\bz_{3\times3}&\bz_{3\times3}&\bz_{3\times1}\\
\bz_{1\times3}&k_a &\bz_{1\times3}&0 \eea\right).\eea\eeq Here
$\texttt{I}_{k\times k}$ and $\bz_{k\times r}$ are the unit and zero
matrices of dimensions ${k\times k}$ and ${k\times r}$ and  $\nu$ is an arbitrary non-zero
parameter.

Thus there exist four equations (\ref{equ}) for spinor and vector
fields which are invariant with respect to indecomposable representations of
the homogeneous Galilei group. The associated matrices $\beta_\mu$ and
$ \beta_4$ are given by formulae (\ref{M1})--(\ref{M4}). All these
equations admit a Lagrangian formulation with the corresponding
Lagrangian of the following standard form
\beq\label{lagra}L=\frac12\psi^\dag\left(\beta_\mu p^\mu
+\beta_4m\right)\psi+h.c..\eeq

The equations (\ref{equ}) with $\beta$-matrices specified in
(\ref{M1}), (\ref{M2}), and (\ref{M4}) are equivalent to ones
analyzed in papers \cite{Hagen} and \cite{nikitin}.
However, at the best of our knowledge equation (\ref{equ}) with
$\beta$-matrices (\ref{M3}) is new. Let us remark that these
matrices satisfy neither relations (\ref{cas10}) nor (\ref{cas11})
and the related equation describes a Galilean quantum mechanical
system whose spin can take two values: $s=1$ and $s=0$. Notice that
matrices $\beta_\texttt{m}$ satisfying (\ref{cas10}) or
(\ref{cas11}) do not exist for representation $D(2,2,1)$.

There exist also a number of  wave equations (\ref{equ}) invariant
with respect to {\it decomposable} representations of $hg(1,3)$.
We shall discuss some of them in detail
in the
sections which follow. A
general description of all such equations is given in  Sections
2.4, 2.5 and in the Appendix.

\section{Galilean analogues of some basic relativistic equations}

All relativistic wave equations have their
Galilei invariant counterparts. For instance, the Galilean analogue of the Dirac
equation is the L\'evy-Leblond equation which was discussed in Section 2.1.

In the present section we consider some basic relativistic wave
equations for particles with higher spin. Using our knowledge of scalar,
spinor and vector representations and of invariants of the homogeneous
Galilei group we construct Galilean forms of these equations.

The Galilean analogues of some basic relativistic equations  form basic non-relativistic equations. They are, however, not obtained by direct non-relativistic limits of basic relativistic equations (but of some other ones).

\subsection{The Galilean second order Proca equation}

The relativistic Proca equation for a vector field $\psi^\mu$ can be written as
\cite{proca}: \beq\label{proca}W^\mu\equiv\left(p_\nu
p^\nu-m^2\right)\psi^\mu-p^\mu p^\nu \psi_\nu=0,\eeq where
$p_\nu=\ri\frac\partial{\partial x^\nu},\ \nu=0,1,2,3,$ and the
raising and lowering of covariant indices $\mu,  \nu$ is made by means of
the relativistic metric tensor
\beq\label{gr}g_{\mu\nu}=diag(1,-1,-1,-1).\eeq

Reducing the l.h.s. of equation (\ref{proca}) by $p_\mu$ we obtain the following
consequence
\begin{equation} \label{div}p_\mu \psi^\mu=0\eeq
and then, putting it into (\ref{proca}),\beq
\label{klein}\left(p_\nu p^\nu-m^2\right)\psi^\mu=0.\eeq
Consequently, the four-vector $\psi^\mu$ satisfies both the Klein-Gordon
equation (\ref{klein}) and the four-divergenceless condition
(\ref{div}). The latest one reduces the number of independent
components of $\psi^\mu$ to 3 as it is required for a wave function
of a spin-one (vector) particle.

Notice that contracting the representation $D(\frac12, \frac12)$
of the Lorentz group to the representation $D_1(1,1,1)$ of the homogeneous
Galilei group
equation (\ref{proca}) can be contracted to the direct sum of
Schr\"odinger equations for $\psi^1, \psi^2, \psi^3$ while the zero
component of $\psi$ be expressed as
\begin{gather}\label{sch}
\psi^0=-\frac1m{p_a}\psi^a\end{gather} Such Galilean system is
completely decoupled with respect to physical components $\psi^1,
\psi^2, \psi^3$ and so is not too interesting.

To construct a consistent Galilean analogue of the Proca equation we
start with a wave function which transforms as a
five-vector with respect to the Galilei group transformations.
Namely, the Galilean Proca equation is found to be of
the following form \beq\label{gproca1}\bea{l}W^\texttt{m}\equiv
p_\texttt{n} p^\texttt{n} \hat\psi^\texttt{m} - p^ \texttt{m}
p_\texttt{n} \hat\psi^\texttt{n}+\lambda \delta^{\texttt{m}\ 0}
m\hat\psi^4=0,\ \
\texttt{m,n} =0,1,2,3,4, \eea\eeq where $\lambda$ is an arbitrary
parameter and $\delta^{\texttt{m} 0}$ is a Kronecker symbol.

Equation (\ref{gproca1}) has the following features distinct from those of
(\ref{proca}):
\begin{itemize}
\item the indices $\texttt{m}$ and $\texttt{n}$ take the values $0,1,2, 3,4$ while in
(\ref{proca}) the indices $\mu$ and $\nu$ run from 0 to 3;
\item the relativistic four-gradient $p_\nu$ is replaced by the Galilean
five-vector $p$ (\ref{p}) whose transformation properties are given
by equation (\ref{pcomp});
\item
raising and lowering of covariant indices $\texttt{m},  \texttt{n}$ is made by using
the Galilean metric tensor (\ref{g})
 instead of (\ref{gr}), thus for example
$p_\texttt{m} \hat\psi^\texttt{m}=p^4\hat\psi^0+p^0\hat\psi^4-
p^1\hat\psi^1-p^2\hat\psi^2-p^3\hat\psi^3$, and
$p_\texttt{n}p^\texttt{n}=2p^0p^4-(p^1)^2-(p^2)^2-(p^3)^2=2mp^0-{\bf
p}^2$;
\item equation (\ref{gproca1}) is invariant w.r.t. the  Galilei
transformations (\ref{11}) provided the wave function $\hat\psi^\texttt{m}$
 cotransforms as a Galilean five-vector, i.e., $
\hat\psi^4\to \exp(if(t,{\bf x}))\hat\psi^4,\\
\hat\psi^a\to \exp(if(t,{\bf x}))(\hat\psi^a+v^a\hat\psi^4), \
\hat\psi^0\to \exp(if(t,{\bf x}))(\hat\psi^0+v^a\hat\psi^a+
\frac{v^2}2\hat\psi^4)$. In other words $\hat\psi^\texttt{m}$ should
transforms as a vector from the carrier space of representation $D(1,2,1)$ of
the homogeneous Galilei group.
\end{itemize}

Equation (\ref{gproca1}) admits, like (\ref{proca}), a Lagrangian
formulation and describes a particle with spin 1.
The corresponding Lagrangian has the following form:
\beq\label{lagran}L=(p_\texttt{m}\hat\psi_\texttt{n}-
p_\texttt{n}\hat\psi_\texttt{m})^*
(p^\texttt{m}\hat\psi^\texttt{n}-p^\texttt{n}\hat\psi^\texttt{m})-(p_\texttt{m}
 \hat\psi_\texttt{n})^*
p^\texttt{m} \hat\psi^\texttt{n}+
(p^\texttt{m}  \hat\psi_\texttt{m})^*p_\texttt{n}
\hat\psi^\texttt{n}-\lambda m^2\hat\psi_0^* \hat\psi^4,\eeq
where $p^\texttt{m}$ are components of five-vector $p$ (\ref{p}) and the asterisk
is used to denote complex conjugation.

The system of equations
(\ref{gproca1}) is coupled and can be used to describe spin effects if we
introduce a minimal interaction with an external field, see Section 4.2. In
absence of interaction it is equivalent to Schr\"odinger equations for
three vector components of $\psi^\texttt{m}$. Indeed,
reducing five-vector $W^\texttt{m}$ (whose components are given by
equation (\ref{gproca1})) by $p_\texttt{m}$ we immediately obtain the consequence:
 $\lambda m^2 \hat\psi^4=0$, i.e.,  $\hat\psi^4=0$.
Then considering equation (\ref{gproca1}) for $\texttt{m}=4$ we
conclude that $mp_\texttt{n} \hat\psi^\texttt{n}=0$ and equation
(\ref{gproca1}) is reduced to the following system
\beq\label{free}\bea{l}(2mp_0-{\bf  p}^2)
\hat\psi^\texttt{m}=0,\\m\hat\psi^0+p_a
\hat\psi^a=0,\ \texttt{and } \hat\psi^4=0,\eea\eeq where $ a=1,2,3$.

In accordance with (\ref{free}) the  wave function
$\hat\psi^\texttt{m}$ satisfies the Schr\"odinger equation
componentwise and  has three non-zero components in the rest frame,
which transform as a three-vector under rotations. Consequently equation
(\ref{gproca1}) describes a Galilean particle with spin $s=1$.

Notice that the system of equations (\ref{gproca1}) includes a
five-component wave function and thus cannot be obtained as a simple
non-relativistic approximation of the Proca equation. This is in
accordance with the fact that some of indecomposable representations
 of the homogeneous Galilei algebra (and representation $D(1,2,1)$
 in particular) cannot be obtained by  contraction of irreducible
 representations of the Lorentz algebra, see \cite{Marc} for more
 details.

\subsection{The Galilean Duffin-Kemmer equation}
The relativistic Duffin-Kemmer equation can be written as \cite{kemmer}
\beq\label{KD1} (\beta_\mu p^\mu-\kappa)\psi=0.\eeq Here $\beta_\mu\
(\mu=0,1,2,3)$ are square matrices which satisfy the
Duffin-Kemmer-Petiau (DKP) algebra
\beq\label{KD2}\beta_\mu\beta_\nu\beta_\sigma+
\beta_\sigma\beta_\nu\beta_\mu=2g_{\mu\nu}\beta_\sigma+
2g_{\sigma\nu}\beta_\mu,\eeq where $g_{\mu\nu} $ is a metric tensor given in
(\ref{gr}).

The ring of DKP matrices has two non-trivial irreducible
representations with matrices of dimension $5\times 5$ and $10\times 10$.
The associated  equations (\ref{KD1}) describe  relativistic particles
with spin 0 and 1 correspondingly.

Considering matrices (\ref{M4}) we conclude that the related matrices
\beq\label{Nied}\tilde \beta_\mu=\eta\beta_\mu, \ \mu=0,1,2,3, \
\texttt{ and  } \tilde\beta_4=\eta\beta_4-\nu
\texttt{I}_{10\times10},\eeq where $\eta$ is an invertible matrix
\beq\label{eta}\eta=\left(\bea{cccc}\bz_{3\times3}&\bz_{3\times3}&
\texttt{I}_{3\times3}&\bz_{3\times1}\\
\bz_{3\times3}&\texttt{I}_{3\times3}&\bz_{3\times3}&\bz_{3\times1}\\
\texttt{I}_{3\times3}&\bz_{3\times3}&\bz_{3\times3}&\bz_{3\times1}\\
\bz_{1\times3}&\bz_{1\times3}&\bz_{1\times3}&-1\eea\right)\eeq
 satisfying relations (\ref{KD2}), where, however,
$\mu,\nu,\sigma=0,1,2,3,4$ and $g_{\mu\nu}$ has "Galilean form"
(\ref{g}). Following \cite{fer} we say that such relations define a
Galilean DKP algebra.

Equation (\ref{equ}) with matrices $\tilde\beta_\mu, \tilde\beta_4$ which
satisfy the Galilean DKP algebra
is called the Galilean Duffin-Kemmer equation.
Notice that this equation coincides with equation (\ref{equ}) multiplied
by the invertible matrix $\eta$ provided matrices $\beta_\mu, \ \beta_4$
have the form (\ref{M4}). For the first time the Galilean Duffin-Kemmer equation was considered in \cite{nikitin}.

The Galilean Duffin-Kemmer equation for spin zero particle can also  be
written in the form (\ref{equ}) where $\beta_\mu$ and
$\beta_4$ are $6\times 6$
matrices satisfying a Galilean DKP algebra.
These matrices can be chosen  up to equivalence in
the form \beq\label{KD3}\bea{l}\beta^0=-e_{56}-e_{61},\ \
\ \beta^a=e_{6,1+a}-e_{1+a,6},\ \  a=1,2,3,
\\
\beta^4= e_{15}-e_{22}-e_{33}-e_{44}+e_{51}+ e_{66}
+e_{16}+e_{65},\eea\eeq where $e_{ab}$
denotes a $6\times 6$-dimensional matrix with $1$ at the $ab$ entry, and
$0$ everywhere else.

Notice that matrices (\ref{KD3}) multiplied by the following hermitizing
matrix
\[\hat \eta= e_{15}-e_{22}-e_{33}-e_{44}+e_{51}+ e_{66}\]
became a particular case of our general solution for $\beta$--matrices
(presented in
equations (\ref{b1}), (\ref{b3}), (\ref{b4}) and in the Appendix) for
a direct sum of representations $D(1,2,1)$ and $D(0,0,0)$.

Let us return to the relativistic Duffin-Kemmer equation (\ref{KD1}) and
contract it directly to non-relativistic (i.e., Galilei invariant)
approximation. It is convenient to start with the following tensorial form of this
equation \cite{proca}
\beq\label{KD4}\bea{l}p^\mu\Psi^\nu-p^\nu\Psi^\mu=\kappa\Psi^{\mu\nu},\\
p_\nu\Psi^{\nu\mu}=\kappa\Psi^\mu\eea\eeq
where $\Psi^\mu$ and $\Psi^{\mu\nu}$ are four-vector and skew-symmetric
spinor which transform according to the representation $D(\frac 12,
\frac 12)\oplus D(1,0)\oplus D(0,1)$ of the Lorentz group.

It was shown in papers \cite{Marc} and \cite{NN2} how this representation
can be contracted to representation $D_1(3,1,1)$ of the homogeneous Galilei
group. This contraction can be used to reduce equation (\ref{KD4})
to a Galilei invariant form.
To do this it is necessary:\begin{itemize}
 \item To choose the following new dependent variables
\[\bea{l}R^a=-\frac12(\Psi^{0a}+\Psi^a),\ \ N^a=\Psi^{0a}-\Psi^a,\ \
W^c=\frac12\varepsilon^{abc}\Psi_{bc}, \ \ B=\Psi^0,\eea\]
which, in accordance with (\ref{KD4}), satisfy the following equations:
\beq\label{eququ}\bea{l}2(p^0-\kappa )R^a+p^aB+\varepsilon^{abc}p_bW_c=0,\\
(p^0+\kappa
)N^a-\varepsilon^{abc}p_bW_c+p^aB=0,\\\varepsilon^{abc}p_b(R_c+
\frac12N_c)=\kappa W^a,\\\frac12p_aN^a-p_aR^a=\kappa B;\eea\eeq
\item To act on variables $R^a, N^a, W^a$ and $B$ by a diagonal
contraction matrix. This action yields the change \[R^a=\tilde R^a, \
N^a=\varepsilon^2 \tilde N^a, \ W^a=\varepsilon \tilde W^a, \
B=\varepsilon\tilde B\] where $\varepsilon$ is a small parameter
associated with the inverse speed of light;
\item To change relativistic four-momentum $p^\mu$ and mass $\kappa$ by their
Galilean counterparts  $\tilde p^a, \tilde p^0$ and $m$
where\beq\label{OO} \tilde p^a=\varepsilon^{-1} p^a$, $\tilde
p^0=p_0-\kappa$ and $m=\frac12(p_0+\kappa)\varepsilon^{-2};\eeq
\item In each equation in (\ref{eququ}) keep only
terms which are multiplied by lowest present powers of $\varepsilon$.
\end{itemize}

As a result we obtain the following equations for $\tilde R^a, \tilde N^a,
\tilde W^a$ and $\tilde B$:
\begin{gather}\label{equq}\bea{l}2\tilde p^0 \tilde R^a+\tilde p^a\tilde B+
\varepsilon^{abc}\tilde p_b\tilde W_c=0,
\\\varepsilon^{abc}\tilde p_b\tilde R_c
=m\tilde W^a,\\\tilde p_a\tilde R^a+m\tilde B=0,\eea\\
\label{aquq}2m\tilde N^a=\varepsilon^{abc}\tilde p_b\tilde W_c-
\tilde p^a\tilde B.\end{gather}

The system (\ref{equq}) is nothing else but the Galilei-invariant
equation (\ref{equ}) with matrices (\ref{M2}) written componentwise.
Relation (\ref{aquq}) expresses the extra component $\tilde N^a$ via
derivatives of the essential ones, i.e. of $W^a$ and $B$.

Thus the Galilean analogue of the Duffin-Kemmer equation
(\ref{equ}),
(\ref{M4}) cannot be obtained as a
non-relativistic limit  of  the relativistic Duffin-Kemmer equations
(\ref{KD1}), (\ref{KD2}) but is a specific generalization of it. The relativistic counterpart of equations
(\ref{equ}), (\ref{M4}) is a specific generalization of (\ref{KD1}),
(\ref{KD2}) which will be studied in a separate publication.

\subsection{The Galilean Rarita-Schwinger equation}

Till now we have used our knowledge of indecomposable representations of algebra
$hg(1,3)$ for spinor, scalar and vector fields to construct wave equations
for
fields of spin $\tilde s\leq1$. In this section we derive Galilean
invariant equations for the field transforming as a direct product of
spin 1/2 and spin 1 fields. The relativistic analogue of such system
is the famous Rarita-Schwinger equation.

The relativistic Rarita-Schwinger equation for a particle with spin
$s=\frac32$ is constructed by using a vector-spinor wave function
$\Psi^\mu_\alpha$, where $\mu=0,1,2,3 $ and $\alpha=1,2,3,4$ are
vector and spinor indices respectively. Moreover,
$\Psi^\mu_\alpha$ is supposed to satisfy the equation
\beq\label{rs1}\left(\gamma^\nu p_\nu-m\right)\Psi^\mu-\gamma^\mu p_\nu
\Psi^\nu-p^\mu\gamma_\nu\Psi^\nu
+\gamma^\mu\left(\gamma_\nu p^\nu+m\right)\gamma_\sigma\Psi^\sigma=0,\eeq
where
$\gamma^\mu$ are the Dirac matrices acting on the spinor index $\alpha$
of $\Psi^\mu_\alpha$ which we do not write explicitly.

Reducing the left hand side of equation (\ref{rs1}) by $p_\mu$ and
$\gamma_\mu$ we obtain the following expressions:
\beq\label{rs2}\gamma_\mu\Psi^\mu=0,\ \texttt{and } p_\mu\Psi^\mu=0,\eeq
which reduce the number of independent components of $\Psi^\mu_\alpha$
to 8 as required for a wave function of a relativistic particle with
spin 3/2.

Using our knowledge of invariants for Galilean vector fields from
\cite{Marc},  \cite{NN2}  we can easily find a Galilean analogue of equation
(\ref{rs1}). Like in the case of the Galilean Proca equation
we begin with
a five-vector $\hat\Psi^\texttt{m},\ \texttt{m}=0,1,2,3,4$  which has, in addition,
a bi-spinorial index which we do not write explicitly.
Thus our Galilei invariant equation can be written in the following form:
\beq\label{ra}\hat\gamma_\texttt{n} p^\texttt{n} \hat\Psi^\texttt{m}-\hat\gamma^
\texttt{m} p_\texttt{n}\hat\Psi^\texttt{n}-p^\texttt{m}\hat\gamma_\texttt{n}\hat\Psi^\texttt{n}+\hat\gamma^\texttt{m}\hat
\gamma_\texttt{n} p^\texttt{n}\hat\gamma_\texttt{r}\hat\Psi^\texttt{r}+
\lambda\delta^{\texttt{m} 0}m\hat\Psi^4=0.\eeq
Here, $\hat\gamma_\texttt{n}$ are the Galilean $\gamma$-matrices
(\ref{gamma}),
$p^\texttt{m}$ is the Galilean "five-momentum"  (\ref{p}),
$\lambda$ is an
arbitrary non-vanishing parameter and
raising and lowering of indices $\texttt{m and }\texttt{n}$ is made by
using the Galilean metric tensor (\ref{g}).

Reducing (\ref{ra}) by $p_\texttt{m}$ we receive
$\lambda m^2\hat\Psi^4=0$, i.e., $\hat\Psi^4=0$. Whereas, reducing
(\ref{ra})
by $\hat\gamma_\texttt{m}$ we obtain
\beq\label{con}p_\texttt{n}\hat\Psi^\texttt{n}=
\hat\gamma_\texttt{n} p^\texttt{n}\hat\gamma_\texttt{m}\hat
\Psi^\texttt{m}.\eeq
Finally, comparing (\ref{con}) with equation (\ref{ra}) for $\texttt{m}=4$
we find the following consequences of equation (\ref{ra}):
\begin{gather}
\label{ra1}\hat\gamma_\texttt{n} p^\texttt{n}\hat\Psi^\sigma=0,\ \ \ \
\sigma=0,1,2,3,\\
\label{ra2}m\hat\Psi^0-p^a\hat\Psi^a=0,\ \ \ a=1,2,3,\\
\label{ra3}\hat\gamma_0\hat\Psi^0+\hat\gamma_a\hat\Psi^a=0,\ \texttt{and }
\hat\Psi^4=0.\end{gather} On the other hand equation
(\ref{ra}) follows from (\ref{ra1})--(\ref{ra3}), so that
equations (\ref{ra}) and (\ref{ra1})--(\ref{ra3}) are equivalent.

In accordance with (\ref{ra1}) any component of $\Psi^\texttt{m}$
satisfies the L\'evy-Leblond equation, compare with Section 2.1.
Let us prove that equations (\ref{ra1})--(\ref{ra3}) describe indeed
a particle with spin s=3/2.

It follows from (\ref{ra2})--(\ref{ra3}) that
$\hat\Psi^0=\hat\Psi^4=0 $ in the rest frame. Using the realization
(\ref{gamma}) for $
\hat\gamma$-matrices we conclude that, in accordance with (\ref{ra3}),
$\hat\Psi^a$ satisfies the equation
\beq\label{rs10}\hat\sigma_a\hat\Psi^a=0,\texttt{ where }
\hat\sigma_a=\texttt{I}_{2\times2}\otimes \sigma_a.\eeq It follows
from
 (\ref{rs10}) that function $\hat\Psi^a$ satisfies
conditions (\ref{cas6}) and (\ref{cas61}) with $s=3/2$ since the
total spin operator $\bf S$ is a sum of operators of spin one and of
spin one--half: \beq\label{rs11}{ S_a}=\hat s_a+\frac12
\hat\sigma_a,\ \hat s_a=\texttt{I}_{2\times2}\otimes s_a.\eeq Hence
\beq\label{rs12}{\bf S}^2=\frac{11}{4}+\hat{\bf
s}\cdot\hat{\mbox{\boldmath$\sigma$\unboldmath}}.\eeq

Let  $\tilde\Psi$ denotes the column $(\hat\Psi^1, \hat\Psi^2,
\hat\Psi^3)$. In accordance with (\ref{rs12}) the condition ${\bf
S}^2\tilde\Psi=s(s+1)\tilde\Psi$ reduces to the form
\beq\label{rs13}\hat\Psi_a-\frac{i}{2}\varepsilon_{abc}\hat
\sigma_b\hat\Psi_c=0,\eeq for $s=3/2$ provided we use the representation
with
$(s_a)_{bc}=i\varepsilon_{abc} $, where $\varepsilon_{abc}$ is a
totally antisymmetric unit tensor.

Comparing (\ref{rs10}) with (\ref{rs13}) we conclude that these equations
are completely equivalent
since multiplying (\ref{rs10}) by $\hat\sigma_a$ we obtain (\ref{rs13})
and multiplying (\ref{rs13}) by $\hat s_a$ and contracting
it with
respect to index $a$ we receive (\ref{rs10}).
Thus indeed equation (\ref{ra}) describes a Galilean particle with spin
$s=3/2$.

Like equation (\ref{gproca1}) the equation (\ref{ra}) admits a Lagrangian
formulation.
The corresponding Lagrangian can be written as
\beq\label{lag}L=\frac12(\bar{\hat\Psi}_\texttt{m} \hat\gamma_\texttt{n}
p^\texttt{n} \hat\Psi^\texttt{m}
-\bar{\hat\Psi}_\texttt{m}\hat\gamma^\texttt{m} p_
\texttt{n}\hat\Psi^\texttt{n}-\bar{\hat\Psi}_\texttt{m}
p^\texttt{m}\hat\gamma_\texttt{n}\hat\Psi^\texttt{n}+
\bar{\hat\Psi}_\texttt{m}\hat\gamma^\texttt{m}\hat\gamma_\texttt{n}
p^\texttt{n}\hat\gamma_\alpha
\hat\Psi^\alpha+
\lambda m\bar{\hat\Psi}_0\hat\Psi^4)+h.c.,\eeq
where $\bar {\hat\Psi}_\texttt{m}=
\hat\Psi_\texttt{m}^\dag\eta$ and $\eta$ is the
hermitizing matrix (\ref{geta}).

We note that it is possible to find a Galilei-invariant equation for
particle with spin $3/2$ starting with the relativistic equation
(\ref{rs1}) and making the In\"on\"u-Wigner contraction of
the representation of the Lorentz group realized on solutions of the
Rarita-Schwinger equation. However the related theory appears to be rather
cumbersome in comparison with our Galilei-invariant Rarita-Schwinger equation (\ref{ra}).

\section{Equations for charged particles interacting with an external
gauge field}

\subsection{Minimal interaction with an external field}

We have described Galilei invariant equations (\ref{equ}) for free
particles with spins 0, 1/2,1 and 3/2. These equations have admitted
Lagrangian formulation (\ref{lagra}), so that to generalize them to the
case of particles interacting with an external field means as usually to
apply the
minimal interaction principle, i.e., to make the following change in
the Lagrangian \beq\label{minimal}p^\mu\to
\pi^\mu=p^\mu-eA^\mu,\ p^4\to \pi^4=p^4-eA^4,\eeq where $A^\mu$  and
$A^4$ are components of a vector-potential of the external field,
$p^4=m$ and $e$ is a particle charge.

Thus we come to the Lagrangian
\beq\label{minimal_equation}L=\frac12\Psi^\dag\left(\beta_\mu
\pi^\mu+\beta_4\pi^4\right) \Psi({\bf x}, t)+h.c. \eeq

It is important to note that change (\ref{minimal}) is
compatible with the Galilei invariance provided the components
$(A^0,A^1,A^2,A^3,A^4)$ of the vector-potential transform as a
Galilean five-vector, i.e., as \beq\label{trans}A^0\to A^0+{\bf v}\cdot
{\bf A}+\frac{{\bf v}^2}2A^4,\ {\bf A}\to{\bf A}+{\bf v}A^4,\ A^4\to
A^4.\eeq Formula (\ref{minimal_equation}) presents the most general
Lagrangian for the Galilean Bhabha equation describing a
particle interacting with an external gauge field via minimal
interaction. On the other hand, in contrast to the relativistic
case, there are many other possibilities. Here we shall mention
several of them concerning various types of Galilean massless
fields.
\begin{itemize}
\item First we consider a four-vector potential $A=(A^0, {\bf A})$ which
transforms according to the representation $D(1,1,1)$, i.e., via the
following expression under Galilei boost:
 \beq A^0\to A^0+{\bf v}\cdot{\bf A}, \ {\bf A}\to{\bf
A}.\label{A^0}\eeq
 Such potential corresponds to a "magnetic" limit of the Maxwell
 equations,
 see \cite{ll1967}, \cite{lebellac}.
\item The second possibility concerns a four-vector potential
$A=({\bf A},\ A^4)$ which transforms according to the representation
$D(1,1,1)$, i.e., \beq A^4\to A^4, \ {\bf A}\to{\bf A}+{\bf
v}A^4.\label{A^4}\eeq It corresponds to an "electric" limit of the
Maxwell equations \cite{lebellac}.
\item The third and fourth cases consists of either a three-vector potential ${\bf A}$ which
transforms according to the representation $D(1,0,0)$ or the scalar
potential $A^4$ which transforms  according to the representation
$D(0,1,0)$. Both these potentials are invariant with respect to
Galilei boosts.
\item The fifth possibility is formed by the above mentioned potentials restricted by some Galilei
invariant conditions. For example it is possible to impose on potential  $A=({\bf A},\ A^4)$ the
condition   $\nabla A^4=0$ .
The complete description of all Galilean vector-potentials will be given
elsewhere.    \end{itemize}

The corresponding equations for a charged particle interacting with
an external field are the Euler equations derived from the
Lagrangian (\ref{minimal_equation}) in which the non-zero components of $A$ specified in the
above Items.

Thus there are many possibilities how to generalize Lagrangians
(\ref{lagra}) for particles interacting with external fields via
minimal interaction. However, it is also possible to introduce
interaction via other means, e.g., via the anomalous (Pauli) term.

\subsection{The Galilean Proca equation for interacting particles}

Let us consider now the Galilean Proca equation for a charged particle
interacting with an external e.m. field in magnetic limit. For this
purpose we introduce a minimal interaction into the Lagrangian
(\ref{lag}), i.e., make the change $p^\texttt{n}\to \pi^\texttt{n}$
therein. The corresponding Euler-Lagrange equation is then of the
form \beq\label{gproca11}\bea{l}\widehat W^\texttt{m}\equiv
\pi_\texttt{n} \pi^\texttt{n} \hat\psi^\texttt{m}+2\ri
eF^{\texttt{mn}}\hat \psi_\texttt{n} - \pi^ \texttt{m}
\pi_\texttt{n} \hat\psi^\texttt{n}+\lambda \delta^{\texttt{m}\ 0}
m\hat\psi^4=0, \eea\eeq where
$eF^{\texttt{mn}}=-i[\pi^\texttt{m},\pi^\texttt{n}]$, so that
\beq\label{gproca12}F^{0a}=-F^{a0}=-E_a, \
F^{ab}={\varepsilon^{ab}}_cH_c,\ F^{4\mu}=-F^{\mu 4}=0.\eeq

 To evaluate equations (\ref{gproca11}) we change  $\hat \psi^\texttt{m}\to\psi^\texttt{m}$, namely:
\beq\label{gproca13}\hat \psi^4=\psi^4,\ \
{\hat{\mbox{\boldmath$\psi$\unboldmath}}}=
{\mbox{\boldmath$\psi$\unboldmath}}+{\mbox{\boldmath$\pi$\unboldmath}}
\psi^4,\ \ \hat
\psi^0=\psi^0+\frac1m{\mbox{\boldmath$\pi$\unboldmath}}
\cdot{\mbox{\boldmath$\psi$\unboldmath}}+
\frac{\mbox{\boldmath$\pi$\unboldmath}^2}{2m^2}\psi^4.\eeq This is
nothing else then the specified later similarity transformation
$\hat \psi\to \psi=W\hat\psi$, where $W$ is operator (\ref{W}) with
the appropriate matrices ${\mbox{\boldmath$\eta$\unboldmath}}$.

 Substituting
(\ref{gproca12}) and (\ref{gproca13}) into (\ref{gproca11}) we
obtain the following system of equations
\begin{gather}\label{gproca14}\bea{l}{\widehat W}^4=\frac12(2m\pi^0-
{\mbox{\boldmath$\pi$\unboldmath}}^2
)\psi^4-m^2\psi^0=0,\eea\\\label{gproca15}\bea{l}\widehat{\bf
W}=(2m\pi^0-
{\mbox{\boldmath$\pi$\unboldmath}}^2)({\mbox{\boldmath$\psi$\unboldmath}}+
\frac1m {\mbox{\boldmath$\pi$\unboldmath}}\psi^4)+2\ri e{\bf
H}\times{\mbox{\boldmath$\psi$\unboldmath}}-(2\ri e{\bf F}+e^2{\bf
j})\psi^4\\-{\mbox{\boldmath$\pi$\unboldmath}}(m\psi^0+\frac12(2m\pi^0-
{\mbox{\boldmath$\pi$\unboldmath}}^2)\psi^4)=0,\eea\\
\label{gproca16}\bea{l}\widehat W^0=(2m\pi^0-
{\mbox{\boldmath$\pi$\unboldmath}}^2)(\psi^0+
\frac1m{\mbox{\boldmath$\pi$\unboldmath}}
\cdot{\mbox{\boldmath$\psi$\unboldmath}}+
\frac{\mbox{\boldmath$\pi$\unboldmath}^2}{2m^2}\psi^4)+\lambda
m^2\psi^4\\-2ie{\bf
E}\cdot({\mbox{\boldmath$\psi$\unboldmath}}+\frac1m
{\mbox{\boldmath$\pi$\unboldmath}}\psi^4)-\pi^0(m\psi^0+
\frac12(2m\pi^0-
{\mbox{\boldmath$\pi$\unboldmath}}^2)\psi^4)=0\eea
\end{gather}
where we use the following notation $\texttt{div} {\bf E}=ej^0$ and
$\texttt{curl}
 {\bf H}=e{\bf j}$.

 In accordance with (\ref{gproca14}) $\psi^0$ can be expressed via
 derivatives of $\psi^4$. In this way the following coupled system
 of equations can be derived from (\ref{gproca15}) and
 (\ref{gproca16}):
\begin{equation}\label{gproca17}\left(\pi^0-
\frac{{\mbox{\boldmath$\pi$\unboldmath}}^2}{2m}+\frac{e}{m}{\bf
s}\cdot{\bf H}\right){\mbox{\boldmath$\psi$\unboldmath}}=
\frac{e^4}{2\lambda m^5}{\bf j}({\bf
j}\cdot{\mbox{\boldmath$\psi$\unboldmath}}-\frac1m(j^0m-{\bf
j}\cdot{\mbox{\boldmath$\pi$\unboldmath}})\psi^4),\end{equation} and
\begin{equation}\label{gproca18}(\lambda m^4+ {e^2}(j^0m-{\bf
j}\cdot{\mbox{\boldmath$\pi$\unboldmath}}))\psi^4= {e^2} m {\bf
j}\cdot{\mbox{\boldmath$\psi$\unboldmath}}.
\end{equation}

The r.h.s. of equation (\ref{gproca17}) includes a very small
multiplier $\frac{e^4}{2\lambda m^5}$ and therefore the wave
function $\bf \psi$ satisfies the Schr\"odinger-Pauli equation with
a very high accuracy. Moreover, a gyromagnetic ratio (i.e., the
coefficient at the term $\frac{e}{2m}{\bf s}\cdot{\bf H})$ has the
desired value which is equal to 2.

\subsection{Galilean Bhabha equations with minimal and anomalous
interactions}

Taking Lagrangian (\ref{minimal_equation}) we can derive the
following equation for a charged particle interacting with an
external field
\beq\label{minimal_equation1}L\Psi\equiv\left(\beta_\mu
\pi^\mu+\beta_4\pi^4\right) \Psi({\bf x}, t)=0. \eeq

Let us consider now equation (\ref{minimal_equation1}) with general
matrices $\beta_\mu$ and $ \beta_4$. If we restrict ourselves to a
vector-potential of magnetic type, i.e., to $A=(A^0,{\bf A},0)$,
then \beq\label{min}\pi^0=p^0-eA^0,\ \pi^a=p^a-eA^a,\ \pi^4=m.\eeq

Like free particle equations (\ref{equ}) the equations
(\ref{minimal_equation1}) are Galilei-invariant provided matrices
$\beta_\mu, \beta_4$ satisfy conditions (\ref{cond}). In addition
the vector-potential of an external field  should transform
according to properties (\ref{A^0}).

Following Pauli \cite{39} we generalize our equation
(\ref{minimal_equation1}) by adding to it an interaction terms
linear in electromagnetic field strengthes and consider the
equation:
\beq\label{anint}\left(\beta_\mu\pi^\mu+\beta_4m+F\right)\psi=0,\  \texttt{where }
\ F=\frac{e}{m}({\bf A}\cdot {\bf H}+{\bf G}\cdot{\bf E}).\eeq Here,
${\bf A}$ and ${\bf G}$ are matrices determined by requirement
of Galilei invariance, i.e., by the conditions that ${\bf A}\cdot {\bf
H}$ and ${\bf G}\cdot {\bf E}$ should be Galilean scalars.

In paper \cite{Marc} we find the most general form of the Pauli
interaction which can be introduced into the L\'evy-Leblond equation
for a particle of spin 1/2 . Finding the general Pauli interaction
for other Galilean particles is a special problem for any equation
previously considered. Here we restrict ourselves to a systematic analysis of the
Pauli terms which are true for {\it any} Galilean Bhabha equation.

First we shall prove the following statement.

{\bf Lemma}. {\it Let $S_a, \eta_a$ be matrices which realize a
representation of algebra $hg(1,3)$, $\Lambda$ be a matrix
satisfying the conditions \beq\label{L}S_a\Lambda=\Lambda S_a,\ \
\eta^\dag_a\Lambda= \Lambda\eta_a\eeq
 and \[E_a=-\frac{\partial A_0}{\partial x_a}-\frac{\partial A_a}
{\partial t}, H_a=\varepsilon_{abc}\frac{\partial A_b}{\partial
x_c}\] be vectors of the electric and magnetic field strength respectively. Then
matrices \beq\label{paul5}F_1=\Lambda({\bf s}\cdot{\bf H} -
{\mbox{\boldmath$\eta$\unboldmath}}\cdot {\bf E})\eeq and
\beq\label{paul6}F_2=
\Lambda{\mbox{\boldmath$\eta$\unboldmath}}\cdot{\bf H}\eeq are
invariant with respect to the Galilei transformations provided the
vector-potential $A$ is transformed in accordance with Galilean law
(\ref{trans})}.

{\bf Proof}. First we note that matrices (\ref{paul5}) and
(\ref{paul6}) are scalars with respect to rotations. Then, starting
with transformation laws (\ref{11}) and (\ref{trans}) we easily find
that under a Galilei boost the vectors $\bf E$ and $\bf H$
co-transform as \beq\label{cotrans}{\bf E}\to {\bf E}-{\bf
v}\times{\bf H},\ \ \ {\bf H}\to {\bf H}.\eeq

On the other hand transformation laws for matrices $\Lambda\bf S$
and ${\Lambda\mbox{\boldmath$\eta$\unboldmath}}$ can be found using
the exponential mapping of boost generators $\bf G$ given in
equation (\ref{cov}): \beq\label{trah}\bea{l}{\bf S}\to \exp(i{\bf
G}^\dag\cdot{\bf v}) \Lambda{\bf S} \exp(-i{\bf G}\cdot{\bf
v})=\Lambda\exp(i{\mbox{\boldmath$\eta\cdot v$ \unboldmath}}){\bf
S}\exp(-i{\mbox{\boldmath$\eta\cdot v$ \unboldmath}})\\={\bf s}+{\bf
v}\times {\mbox{\boldmath$\eta$\unboldmath}},\\
{\mbox{\boldmath$\eta$\unboldmath}} \to \exp(i{\bf G}^\dag\cdot{\bf
v})\Lambda{\mbox{\boldmath$\eta$\unboldmath}} \exp(i{\bf G}\cdot{\bf
v})=\Lambda\exp(i{\mbox{\boldmath$\eta\cdot v$
\unboldmath}}){{\mbox{\boldmath$\eta$
\unboldmath}}}\exp(-i{\mbox{\boldmath$\eta\cdot v$
\unboldmath}})=\Lambda {\mbox{\boldmath$\eta$ \unboldmath}}
.\eea\eeq

One easily verifies that transformations (\ref{cotrans}) and
(\ref{trah}) leave matrices $F_1$ and $F_2$ invariant.Q.I.D

In accordance with the Lemma there are many possibilities how to
generalize equations previously considered to the case of anomalous
interaction. Indeed, for any
Galilean Bhabha equation there are matrices $S_a$, $\eta_a$ and
$\Lambda$ for which conditions (\ref{L}) are satisfied. For example,
we can choose $\Lambda=\beta_0$. In addition, for many cases there
exist a hermitizing matrix $\eta=\Lambda$ satisfying (\ref{L}), see,
e.g., equations (\ref{geta}) and (\ref{eta}). For particular
representations of algebra $hg(1,3)$ there are also other solutions
of equations (\ref{L}).

Thus the Pauli terms for the Galilean Bhabha equations can be chosen in
the form (\ref{paul5}) or (\ref{paul6}) or, more generally, as a
linear combination of both, $F_1$ and $F_2$. As a result we obtain the
following equation,
\beq\label{anint1}\left(\beta_\mu\pi^\mu+\beta_4m+\lambda_1\frac em
\beta_0{\mbox{\boldmath$\eta$ \unboldmath}}\cdot{\bf
H}+\lambda_2\frac em\beta_0 \left({\bf S}\cdot{\bf
H}-{\mbox{\boldmath$\eta$ \unboldmath}}\cdot{\bf
E}\right)\right)\psi=0,\eeq where $\lambda_1$ and $\lambda_2$ are
dimensionless coupling constants.

Let us analyze a physical content of this equation. Since equation
(\ref{anint1}) is reduced to equation (\ref{minimal_equation1}) for
$\lambda_1=\lambda_2=0$, we shall study the cases with anomalous and
with minimal interaction simultaneously by analyzing a more general
equation (\ref{anint1}).

In order to receive the physical content of these equations it is convenient
to apply the transformation \beq\label{trans1}\Psi\to
\Psi'=W^{-1}\Psi, \ L \to L'= W^\dag LW,\eeq where
\beq\label{W}W=\exp\left(-
i\frac{\mbox{\boldmath$\eta\cdot\pi$\unboldmath}}{m} \right)\eeq and
${\mbox{\boldmath$\eta$\unboldmath}}$ is a vector whose components
are the Galilei boost generators (\ref{02}). For the case $e=0$ (or
$A_\mu=0$) the operator $W$ reduces to operator $U$ given in
equation (\ref{W0}), which was used for our analysis of free
particle equations.

 Using relations
(\ref{cond}) and supposing that the nilpotence index of matrices
${\mbox{\boldmath$\eta\cdot\pi$\unboldmath}}$ is less than 4 we come
to the following equation which is equivalent to (\ref{anint1}):
\beq\label{aprox}\bea{l}L'\Psi'\equiv\left\{(\beta_0\left(\pi^0-\frac{
{\mbox{\boldmath$\pi$\unboldmath}}^2}{2m}+
\frac{e}{m}{\mbox{\boldmath$\eta$\unboldmath}}\cdot {\bf
F}\right)-\frac{e}{2m}{\mbox{\boldmath$\beta$\unboldmath}}
\times{\mbox{\boldmath$\eta$\unboldmath}}\cdot{\bf H}+\beta_4m
-\frac{e}{6m^2}\widehat Q_{ab}\frac{\partial H_a}{\partial
x_b}\right.\\\\\left. +\frac em\Lambda\left[\lambda_1
{\mbox{\boldmath$\eta$ \unboldmath}}\cdot{\bf H}+\lambda_2\left(
{\bf S}\cdot{\bf H}-{\mbox{\boldmath$\eta$ \unboldmath}}\cdot{\bf
F}+\frac 1{2m}\tilde Q_{ab}\frac{\partial H_a}{\partial
x_b}\right)\right]\right\}\Psi'=0\eea\eeq where ${\bf E}=-\nabla
A^0-\frac{\partial {\bf A}}{\partial t}$ and ${\bf H}=\nabla\times
{\bf A}$ are vectors of the corresponding electric and magnetic
field strength respectively,
 \beq\label{quadr}\bea{l}{\bf F}={\bf E}+\frac{1}{2m}
 ({\mbox{\boldmath$\pi$\unboldmath}}\times{\bf
H}-{\bf H}\times{\mbox{\boldmath$\pi$\unboldmath}}),\\  \widehat
Q_{ab}=\eta^\dag_a\varepsilon_{bcd}\beta_c\eta_d+
\varepsilon_{bcd}\beta_c\eta_d\eta_a,\ \ \tilde
Q_{ab}=\eta_aS_b+S_b\eta_a.\eea\eeq

Equation (\ref{aprox}) includes the Schr\"odinger terms $(\pi^0-
\frac{ {\mbox{\boldmath$\pi$\unboldmath}}^2}{2m})\Psi'$ and
additional terms which are linear in vectors of the external field
strengthes and their derivatives.

Notice that if the nilpotence index $N$ of matrices
${\mbox{\boldmath$\eta$\unboldmath}}$ is smaller  than 4, the
transformed equation (\ref{aprox}) is completely equivalent to
initial equation (\ref{anint1}). The condition $N<4$ is fulfilled
for all representations of algebra $hg(1,3)$ considered in the
present paper, so this equivalence takes place for all equations for
vector, scalar and spin 1/2 fields studied in the paper.


\subsection{Galilean equation for spinor and vector
fields with interactions}

Let us consider equation (\ref{aprox}) for two particular
realizations  of $\beta$--matrices in more detail. First notice,
that our conclusions from equations
(\ref{minimal_equation1})--(\ref{aprox}) are true in general and in
particular for the L\'evy-Leblond equation, i.e.,  when $\beta_\mu, \beta_4$
 $4\times4$  are matrices determined by relations (\ref{beta}) with
$\omega=\kappa=0$. Then $\beta_0\eta_a=0,\ Q_{ab}=Q'_{ab}=0, \
{\mbox{\boldmath$\beta$\unboldmath}}
\times{\mbox{\boldmath$\eta$\unboldmath}}=-2\beta_0{\bf S}$, and
equation (\ref{aprox}) is reduced to the following form:
\beq\label{aprox1}\bea{l}\left\{\beta_0\left(\pi^0-\frac{
{\mbox{\boldmath$\pi$\unboldmath}}^2}{2m}+\frac{e}{m}{\bf
S}\cdot{\bf H}\right)+\beta_4m \right.\\\left. +\frac
em\Lambda\left[\lambda_1 {\mbox{\boldmath$\eta$
\unboldmath}}\cdot{\bf H}+\lambda_2\left( {\bf S}\cdot{\bf
H}-{\mbox{\boldmath$\eta$ \unboldmath}}\cdot{\bf
F}\right)\right]\right\}\Psi'=0\eea\eeq

For $\lambda_1=\lambda_2=0$ (i.e., when only the minimal interaction
is present) equation (\ref{aprox1}) is reduced to the following
system:
\beq\label{pauli}\left(\pi_0-\frac{\mbox{\boldmath$\pi$\unboldmath}^2}
{2m}+\frac{e}{2m}{\mbox{\boldmath$\sigma$\unboldmath}}\cdot {\bf
H}\right) \varphi_1=0,\eeq \beq\label{var} m\varphi_2=0, \ \
\texttt{or }\varphi_2=0,\eeq where $\varphi_1=\beta_0\Psi'$ and
$\varphi_2=(1-\beta_0)\Psi'$ are two--component spinors.

Thus introducing the minimal interaction (\ref{minimal}) into the
L\'evy-Leblond equation, we receive the Pauli equation for physical
components of the wave function; moreover, the coupling constant
(gyromagnetic ratio) for the Pauli interaction $\frac{e}{2m}\hat{\bf
s}\cdot {\bf H}$, $\hat{\bf
s}=\frac12{\mbox{\boldmath$\sigma$\unboldmath}}$
 has the same value $g=2$ as in the case of the
Dirac equation \cite{levyleblond}.

Considering now the case with an anomalous interaction we conclude that
the general form of matrix  $\Lambda $ satisfying relations
(\ref{L}) is \beq\label{n7}\Lambda=\nu\beta_0+\mu\eta,\eeq where
$\eta$ is hermitizing matrix (\ref{geta}), $\nu$ and $\mu$ are
arbitrary parameters. Substituting (\ref{n7}) into (\ref{aprox1}) we
obtain the following analogue of system (\ref{pauli}):
\beq\label{paulii}\bea{l}\left(\pi_0-\frac{\mbox{\boldmath$\pi$\unboldmath}^2}
{2m}+\frac{eg}{2m}{\mbox{\boldmath$\sigma$\unboldmath}}\cdot {\bf
H}-\frac{e\lambda_3}{2m}{\mbox{\boldmath$\sigma$\unboldmath}}\cdot
{\bf F}-\frac{\lambda_3^2e^2}{8m^3}{\bf H}^2\right)
\varphi_1=0,\\\varphi_2=-\frac{\lambda_3e}{4m}
{\mbox{\boldmath$\sigma$\unboldmath}}\cdot {\bf H}\varphi_1,\eea\eeq
where $g=2+\mu\lambda_1+\nu\lambda_2, \ \lambda_3=\mu\lambda_2$ are
arbitrary parameters.

We see that the L\'evy-Leblond equation with minimal and anomalous
interactions reduces to the Schr\"odinger-Pauli-like equation
(\ref{paulii}) for two-component spinor $\varphi_1$, which, however,
includes additional terms linear in strengths of an electric field
and linear and quadratic in strengths of a magnetic field. We shall
discuss them at the end of this section.

Consider now the Galilean Duffin-Kemmer equation for interacting
vector field. The corresponding $\beta$-matrices in(\ref{anint1})
have dimension $10\times10$ and are given explicitly by relations
(\ref{M4}). Then \[\bea{l} {\mbox{\boldmath$\beta$\unboldmath}}
\times{\mbox{\boldmath$\eta$\unboldmath}} \cdot{\bf
H}=-\left(\bea{cccc}\bz_{3\times3}&{\bf s}\cdot{\bf H}&
\bz_{3\times3}&2{\bf k}^\dag\cdot{\bf H}
\\{\bf s}
\cdot{\bf H}&\bz_{3\times3}&\bz_{3\times3}&\bz_{3\times1}\\
\bz_{3\times3}&\bz_{3\times3} &\bz_{3\times3}&\bz_{3\times1}\\2{\bf
k}\cdot{\bf H}&\bz_{1\times3}& \bz_{1\times3}&0\eea\right),\eea\]
\beq\label{paul21}\bea{l}
\beta_0{\mbox{\boldmath$\eta$\unboldmath}}=\left(\bea{cccc} {\bf
s}&\bz_{3\times3}&\bz_{3\times3}&\bz_{3\times1}\\\bz_{3\times3}&
\bz_{3\times3}&\bz_{3\times3}&\bz_{3\times1}\\\bz_{3\times3}&
\bz_{3\times3}&\bz_{3\times3}&\bz_{3\times1}\\\bz_{1\times3}&
\bz_{1\times3}&\bz_{1\times3}&0\eea\right),\ \beta_0{\bf S}=
\left(\bea{cccc}\bz_{3\times3}&{\bf s}&\bz_{3\times3}&\bz_{3\times1}
\\{\bf s}&\bz_{3\times3}&\bz_{3\times3}&\bz_{3\times1}\\
\bz_{3\times3}&\bz_{3\times3}&\bz_{3\times3}&\bz_{3\times1}
\\\bz_{1\times3}&\bz_{1\times3}&\bz_{1\times3}&0\eea\right),\\\\\widehat
Q_{ab}=-3\left(\bea{cccc}
Q_{ab}&\bz_{3\times3}&\bz_{3\times3}&\bz_{3\times1}\\\bz_{3\times3}
&\bz_{3\times3}&\bz_{3\times3}&\bz_{3\times1}\\\bz_{3\times3}&\bz_{3\times3}
&\bz_{3\times3}&\bz_{3\times1}\\\bz_{1\times3}&\bz_{1\times3}&\bz_{1\times3}
&0\eea\right),\ \tilde Q_{ab}=\left(\bea{cccc}
Q'_{ab}&\bz_{3\times3}&\bz_{3\times3}&\bz_{3\times1}\\\bz_{3\times3}
&\bz_{3\times3}&\bz_{3\times3}&\bz_{3\times1}\\\bz_{3\times3}&\bz_{3\times3}
&\bz_{3\times3}&\bz_{3\times1}\\\bz_{1\times3}&\bz_{1\times3}&\bz_{1\times3}
&0\eea\right), \eea\eeq where
\beq\label{Q_}Q_{ab}=s_as_b+s_bs_a-\frac43 \delta_{ab},\ \
Q'_{ab}=Q_{ab}+\frac43\delta_{ab}.\eeq Let us consider the
corresponding equation (\ref{aprox}) and restrict ourselves to
$\Lambda=\beta_0$. Representing $\Psi'$ as a column vector
$(\psi_1,\psi_2,\psi_3,\varphi),$ where $\psi_1, \psi_2,\psi_3$ are
three-component vector functions and $\varphi$ is a one-component
scalar function and using (\ref{M4}), (\ref{paul21}) we reduce
(\ref{aprox}) to the following Pauli-type equation for $\psi_1$:
\beq\label{paul3}i\frac{\partial}{\partial t}\psi_1=\widehat
H\psi_1,\eeq where \beq\bea{l}\label{Ha}\widehat H=\frac{\nu^2}{2}m+
\frac{{\mbox{\boldmath$\pi$\unboldmath}}^2}{2m}+ eA_0-\frac {ge}{2
m}{\bf s}\cdot{\bf H}+\frac{qe}{\nu m}{\bf s}\cdot {\bf
E}\\\\-\frac{qe}{2\nu m^2}{\bf
s}\cdot({\mbox{\boldmath$\pi$\unboldmath}}\times {\bf H}-{\bf
H}\times {\mbox{\boldmath$\pi$\unboldmath}})+\frac{e}{2\nu
m^2}(1+\lambda_2)Q_{ab}\frac{\partial H_a}{\partial x_b}+
\frac{e^2}{2\nu^2m^3}\left({\bf H}^2-({\bf s}\cdot{\bf
H})^2\right),\eea \eeq where $g=1+2\lambda_1+2\lambda_2$ and
$q=1-\lambda_2$.

The other components of $\Psi'$ can be expressed by $\psi_1$:
\[\varphi=-\frac{e}{\nu^2m^2}{\bf k}\cdot{\bf H}\psi_1,\
\psi_2=-\frac1\nu\psi_1, \psi_3=-\nu \psi_2-\frac1m \left(\pi_0-
\frac1{2m}{\mbox{\boldmath$\pi$\unboldmath}}^2+ \frac{e}{2m}{\bf
s}\cdot{\bf H}\right)\psi_1.\]

Up to realizations of spin matrices, equation (\ref{paul3}) is
rather similar to equation (\ref{paulii}) describing Galilean
particles with spin 1/2. However, in comparison with (\ref{paulii})the essentially new feature of
equation (\ref{paul3}) appears, namely, that
setting $\lambda_1=\lambda_2=0$ in (\ref{Ha}) (i.e., excluding the
anomalous interaction) we receive a Hamiltonian which still includes
the term $-\frac{e}{\nu m}{\bf s}\cdot{\bf E}$ describing the
coupling of spin with an electric field. We show in the next section
that this effectively represents the spin-orbit coupling.
 The
other terms of Hamiltonian (\ref{Ha}) (which are placed in the
second line of equation (\ref{Ha})) can be neglected starting with a
reasonable assumption about the possible values of the magnetic
field strength.

 However, equation (\ref{paulii}) for $\lambda_1=\lambda_2=0$ it
reduces to the Schr\"odinger-Pauli equation (\ref{pauli}) which has
nothing to do with the spin-orbit coupling.

\subsection{Galilei invariance and spin-orbit coupling}

Consider now the first of equations (\ref{paulii}) for particular
values of arbitrary parameters, namely for $\tilde \lambda_1+\tilde
\lambda_2=-1$: \beq\label{paulik}\hat
L\varphi_1\equiv\left(\pi_0-\frac{\mbox{\boldmath$\pi$\unboldmath}^2}
{2m}-\frac{e\lambda_3}{2m}{\mbox{\boldmath$\sigma$\unboldmath}}\cdot
{\bf F}-\frac{\lambda_3^2e^2}{8m^3}{\bf H}^2\right) \varphi_1=0.\eeq

First, let us remind that this equation is a direct consequence of
the Galilei invariant L\'evy-Leblond equation with anomalous
interaction, i.e., of equation (\ref{anint1}) where
$\beta_\texttt{n}$ are matrices (\ref{beta}) with $\kappa=\omega=0$.
Secondly, equation (\ref{paulik}) by itself is transparently
Galilei invariant since the operator $\hat L$ in (\ref{paulik}) is a Galilean
scalar provided the value of arbitrary
parameter $\lambda_3$ is finite. We shall assume $\lambda_3$ to be small.

In order to find out the physical content of equation (\ref{paulik}) we
transform it to a more transparent
 form using the operator $U=\exp(-\frac{\ri\lambda_3}{2m}
 {\mbox{\boldmath$\sigma\cdot\pi$\unboldmath}})$. Applying this operator
 to $\varphi_1$ and transforming $\hat L\to\hat L'=U\hat LU^{-1}$ we
 obtain the equation
\beq\label{orb}L'\varphi'_1=\left(\pi_0-\frac{{\mbox{\boldmath$\pi$
\unboldmath}}^2}{2m}
-eA_0- \frac{e\lambda_3^2} {8m^2}
\left({\mbox{\boldmath$\sigma$\unboldmath}}\cdot
({\mbox{\boldmath$\pi$\unboldmath}}\times{\bf E}-{\bf
E}\times{\mbox{\boldmath$\pi$\unboldmath}})-\texttt{div} {\bf
E}\right)+\cdots\right)\varphi'_1=0\eeq where the dots denote the
small terms of orders $o(\lambda_3^3)$ and $o(e^2)$.

All terms in square brackets have the exact physical meaning.  They include first
the Schr\"odinger terms
$\pi_0-\frac{{\mbox{\boldmath$\pi$\unboldmath}}^2}{2m} -eA_0$ then the
the term $\sim{\mbox{\boldmath$s$\unboldmath}}\cdot
({\mbox{\boldmath$\pi$\unboldmath}}\times{\bf E}-{\bf
E}\times{\mbox{\boldmath$\pi$\unboldmath}})$ and $\sim
\texttt{div}{\bf E}$ describing the spin-orbit coupling and finally the term $\sim
\texttt{div}{\bf E}$, i.e., the Darwin coupling.

Similarly, starting with equation (\ref{paul3}), setting
$\lambda_2=-1,\ \lambda_1=\frac12$, supposing $\frac1\nu$ to be a small
parameter and making use of a transformation $\psi_1\to\psi'_1=\hat U\psi_1$
with $\hat U=\exp(-\frac{2\ri}{\nu m}{\bf s}\cdot
{\mbox{\boldmath$\pi$\unboldmath}})$ we obtain the
equation
\beq\label{orb1}\bea{l}\left(\pi_0-\frac{\nu^2}2m-
\frac{{\mbox{\boldmath$\pi$\unboldmath}}^2}{2m}
-eA_0\right.\\\left.- \frac{2e} {\nu^2m^2}
\left({\mbox{\boldmath$s$\unboldmath}}\cdot
({\mbox{\boldmath$\pi$\unboldmath}}\times{\bf E}-{\bf
E}\times{\mbox{\boldmath$\pi$\unboldmath}})-Q_{ab}\frac{\partial
E_a}{\partial x_b}+\frac43\texttt{div} {\bf
E}\right)+\cdots\right)\psi'_1=0.\eea\eeq Here the dots denote
small terms of the orders $o(\frac1{\nu^3})$ and $o(e^2)$.

Like relation (\ref{orb}), the equation (\ref{orb1}) includes the terms
which represent the spin-orbit and Darwin
couplings. In addition, there is the term $\sim Q_{ab}\frac{\partial
E_a}{\partial x_b}$ which describes the quadrupole interaction of a
charged vector particle with an electric field.

Thus we again come to the conclusion \cite{nikitin} that the
spin-orbit and Darwin couplings can be effectively described within frame work
of a Galilei-invariant approach and so they have not be
necessarily interpreted as pure relativistic effects.

Let us note that it is possible to choose parameters
$\lambda_1$ and $\lambda_2$ in Hamiltonian (\ref{Ha}) in such a way
that the anomalous interaction with electric field will not be present.
Namely, we can set $\lambda_2=1, \lambda_1=-1/2$, and obtain, instead of (\ref{orb1}), the following
equation :
\beq\label{orb2}\bea{l}\left(\pi_0-\frac{\nu^2}2m-
\frac{{\mbox{\boldmath$\pi$\unboldmath}}^2}{2m}
-eA_0+\frac{ge} {2m}{\bf s}\cdot{\bf H}
+\cdots\right)\psi'_1=0\eea\eeq where $g=2$. In other words,
introducing a specific anomalous interaction into the Galilean Duffin-Kemmer
equation we can reduce it to the Schr\"odinger-Pauli equation with the
correct value of the gyromagnetic ratio $g$.

\section{Discussion}

In the present paper we continue the study of the Galilei invariant
theories for vector and spinor fields, started in \cite{Marc}. The
peculiarity of our approach is that as distinct to the other
approaches (e.g., to \cite{ll1967}-\cite{ourletter}, \cite{santos},
\cite{ijtp}) it enables to find out a complete list
of Galilei invariant equations for scalar and vector fields. This is
possible due to our knowledge of all non-equivalent
indecomposable representations of the Galilei algebra $hg(1,3)$ that
can be
constructed on representation spaces of scalar and vector fields
described for the first time in paper
\cite{Marc}.

Thus using this complete list of representations
we have been able to find all possible systems of the first order Galilei-invariant
wave equations
(\ref{equ}) for scalar and vector fields. All $\beta$-matrices for these
Galilei-invariant wave equations have been presented in the Appendix.
In fact we
have described how to construct arbitrary wave
equations of finite order invariant with respect to the Galilei group since all of
them can be obtained from the first order equations in which various
derivatives of fields are considered
as new dependent variables.

Then Galilean analogues of some popular relativistic
equations for vector particles and particles with spin 3/2 are discussed,
in
particular, the Galilean second order Proca equation and  Galilean first order
Rarita-Schwinger equation. However these Galilean equations are not
non-relativistic limits of the corresponding
relativistic equations Proca and Rarita-Schwinger equation since among other
things they have more components. Thanks to that
it is possible to obtain equations which keep all the main features
of their relativistic analogues.
At the best of our
knowledge this is done for the first time in the present paper.

We pay a specific attention to description of the Galilean particles
interacting with an external electromagnetic field. We
study both the minimal interaction as well as anomalous one. The results
presented in Sections 4.1 and 4.3 are valid for generic equations
describing scalar, spinor or vector fields.

The equation (\ref{anint1}) includes a quite general form of an anomalous
interaction which satisfies the Galilei invariance condition. The
main idea of our analysis of this equation is to transform it to the
equivalent form (\ref{aprox}) in which all terms in brackets commute
with matrix $\beta_0$. The related transformation (\ref{trans1}) can
be treated as a Galilean analogue of the Foldy-Wouthuysen
transformation \cite{foldy}.

Notice that the results presented in Sections 4.1 and 4.3 are valid
for arbitrary equations (\ref{anint1}) invariant with respect to the
Galilei group. The results presented in Sections 4.2, 4.4 and 4.5 are
restricted to particular equations with
anomalous interaction, i.e., to the Galilean Proca
equation, generalized L\'evy-Leblond equation and to generalized Galilean Duffin-Kemmer equation.
We prove that the last two equations
present consistent models of charged particles interacting with an electromagnetic field.
In particular, they
describe such important physical effect as the spin-orbit coupling
which traditionally is interpreted as a pure relativistic phenomenon.

 However, it is
necessary to fix some difficulties of principal nature appearing in
the Galilean approach, like that the Galilei invariance requires that the mass and energy each are separately concerved, and that within the Galilean theories there are not concept of proper time which produces phase effects which do not depend on the velocity of light and so do not dissapear in the non-relativistic limit.
Of course, there are obvious restrictions to
phenomena which are characterized by velocities much smaller than
the velocity of light. In addition, in our approach there are also problems with interpretation of
undesired terms $\sim Q_{ab}\frac{\partial H_a}{\partial x_b}$  and
${\bf s}\cdot({\mbox{\boldmath$\pi$\unboldmath}}\times {\bf H}-{\bf
H}\times {\mbox{\boldmath$\pi$\unboldmath}})$ which appear in the
Hamiltonian (\ref{paul3}). Thanks to appropriate choice of arbitrary
parameters $\lambda_1$ and $\lambda_2$ these terms are not present in
the effective Hamiltonians (\ref{orb1}) and (\ref{orb2}) which
describe spin-orbit and Pauli couplings respectively. However, if we
would like to keep both these couplings, then the undesired terms
may appear.

One more problem is connected with the signes for the terms presenting the spin-orbit and Darwin couplings. Comparing (\ref{orb}) with the quasirelativistic approximation of the Dirac equation (see, e.g., ref. \cite{foldy}) we conclude, that to obtain the correct signes it is necessary to suppose that $\lambda_3$ be purely imaginary. Moreover, for $\lambda_3=\ri$ the coupling constants for spin-orbit and Darwin interactions 
in (\ref{orb}) coinside with the relativistic ones predicted by the Dirac equation. 

Notice that the exact equation (\ref{paulik}) is much simplier then the approximate equation (\ref{orb}) and can be solved exactly for some particular (e.g., the Coulomb) external fields. However, for $\lambda_3$ imaginary  
the term $-\frac{e\lambda_3}{2m}{\mbox{\boldmath$\sigma$\unboldmath}}\cdot
{\bf F}$ in equation (\ref{paulik}) and the corresponding Hamiltonian 
$\hat H=-L+p_0=A_0$ are non-hermitean. On the other hand, for $A_0$ and $\bf A$ being even and odd functions of $\bf x$ correspondingly the equation (\ref{paulik}) appears to be invariant with respect to the product 
of the space inversion $P$ and Wigner time inversion $T$, and so presents one more (and as it is seemd for us, promissing) field for appplication of tools of $PT$-symmetric quantum mechanics  \cite{bender}.

Any Galilean theory by definition is only an approximation of a
relativistic one. The very existence of a physically consistent
non-relativistic approximation can serve as a criteria of
consistency of a relativistic theory. Thus our study of Galilean
wave equations makes a contribution into the theory of relativistic
ones, since effectively we have analyzed possible non-relativistic
limits of theories for vector and scalar particles.

An intriguing problem is the description of Galilean theories for massless
vector and scalar fields. It appears that the fundamental analysis of
Galilean
limits of Maxwell's equations presented in papers \cite{ll1967},
 and \cite{levyleblond} can be essentially completed using the
 list of indecomposable
 representations of algebra $hg(1,3)$ presented in \cite{Marc}.
 This work is in progress.

\section{Appendix}

Here we present all non-trivial solutions of equations (\ref{b2})
which give rise to explicit descriptions of matrices $\beta_4$ given
by equation (\ref{b1}). The related matrices $\beta_0$ and $\beta_a$
are given by equations (\ref{b3}) and (\ref{b4}).

Solving equations (\ref{b2}), where $A$, $C$ and $A'$, $C'$ are
matrices given in Table 1 which correspond to $q=(n,m,\lambda)$ and
$q'=(n', m', \lambda')$ respectively we obtain the related matrices $R=R(q,q'),
E= E(q,q')$ (which define matrix $\beta_4$ in accordance with
equation (\ref{b1})) in the forms presented in Tables 2, where the
Greek letters denote arbitrary real parameters.


\begin{center}

Table 3. Submatrices $R$ and $E$ of matrices $\beta_4$
\end{center}

 \begin{tabular}{|c|c|c|c|}
\hline
& \multicolumn{3}{|c|}{$m,n,\lambda$}\\
\cline{2-4} \raisebox{1.8ex}[0pt][0pt]{$m',n',\lambda'$}
&2,1,1&2,0,0&1,2,1
\\
\hline
2,1,1&$\bea{l}R=\left(\bea{cc}\mu&\nu\\\alpha&0\eea\right)\\
E=\sigma\eea$&$\bea{l}R=\left(\bea{cc}\omega&\nu\\
\mu&0\eea\right)\\E\ \texttt{not}
\\\texttt{existing}\eea$&$\bea{l}R=(\mu\ \nu)\\
E=\left(\bea{c}\sigma\\\alpha\eea\right)\eea$\\\hline
2,0,0&$\bea{l}R=\left(\bea{cc}\mu&\nu\\
\omega&0\eea\right)\\E\ \texttt{not}
\\\texttt{existing}\eea$&$\bea{l}R=\left(\bea{cc}\mu&\nu\\\nu&0\eea\right)\\
E\ \texttt{not} \\\texttt{
existing}\eea$&$\bea{l}R=\left(\bea{cc}\mu&\nu\eea\right)\\
E\ \texttt{not}  \\\texttt{existing}\eea$\\\hline
1,2,1&$\bea{l}R=\left(\bea{c}\mu\\\nu\eea\right)\\
E=(\sigma \ \ \alpha)\eea$&$\bea{l}R=\left(\bea{c}\mu\\\nu\eea\right)\\
E\ \texttt{not}  \\\texttt{existing}\eea$&$\bea{l}R=\mu\\
E=\left(\bea{cc}\mu&\nu\\\nu&0\eea\right)\eea$\\\hline
1,1,0&$\bea{l}R=\left(\bea{c}\mu\\\nu\eea\right),\
E=\sigma\eea$&$\bea{l}R=\mu,\ \
 E\ \texttt{not} \\\texttt{
existing}\eea$&$\bea{l}R=\mu\\
E=\left(\bea{c}\nu\\0\eea\right)\eea$\\\hline
1,1,1&$\bea{l}R=\left(\bea{c}\mu\\\nu\eea\right), \
E=\sigma\eea$&$\bea{l}R=\mu,\ \
 E\ \texttt{not} \\\texttt{
existing}\eea$&$\bea{l} R=\mu\\
E=\left(\bea{c}\nu\\\alpha\eea\right)\eea$\\\hline
1,0,0&$\bea{l}R=\left(\bea{c}\kappa\\\sigma\eea\right)\\
E\ \texttt{not} \\\texttt{ existing}\eea$&$\bea{l}R=\mu,\ \
 E\ \texttt{not} \\\texttt{
existing}\eea$&$\bea{l}R=\mu,\\
 E\ \texttt{not} \texttt{
existing}\eea$\\\hline 0,1,0&$\bea{l}R\ \texttt{not existing},\\
E=\alpha\eea $&$\bea{l}R \texttt{ and }E\\ \texttt{not
existing}\eea$&$\bea{l} R \ \texttt{not existing}\\
E=\mu\eea$\\\hline

\end{tabular}

\newpage

\begin{center}

Table 2. Submatrices $R$ and $E$ of matrices $\beta_4$

\end{center}

\begin{tabular}{|c|c|c|c|}
\hline
& \multicolumn{3}{|c|}{$m,n,\lambda$}\\
\cline{2-4} \raisebox{1.8ex}[0pt][0pt]{$m',n',\lambda'$}
&3,1,1&2,2,1&
2,1,0\\
\hline
3,1,1&$\bea{l}R=\left(\bea{ccc}\mu&\nu&\sigma\\\nu&\alpha&1\\\sigma&1&0\eea\right)
\\
E=\alpha-2\sigma\eea$&
$\bea{c}R=\left(\bea{ccc}\nu&\alpha&0\\\mu&\sigma&\omega\eea\right)\\
E=\left(\bea{c}\kappa\\\omega-\alpha\eea\right)\eea$&$\bea{c}
R=\left(\bea{ccc}\nu&\alpha&\omega\\\mu&\sigma&0\eea\right)\\\\
E=\bea{c}\kappa\eea\eea$\\\hline
2,2,1&$\bea{l}R=\left(\bea{cc}\mu&\nu\\\sigma&\alpha\\\omega&0\eea\right)\\
E=(\kappa\ \ (\omega-\alpha))\eea$&$\bea{l}R=\left(\bea{cc}\mu&\nu\\
\nu&\kappa\eea\right)\\
E=\left(\bea{cc}\sigma&\omega\\\omega&\kappa\eea\right)\eea$&$\bea{l}R=\left(
\bea{cc}\mu&\sigma\\\nu&\omega\eea\right)\\\\E=(\kappa\
\omega)\eea$\\\hline
2,1,0&$\bea{l}R=\left(\bea{ll}\mu&\nu\\\sigma&\alpha\\0&\omega\eea\right)\\
E=\kappa\eea$&$\bea{l}R=\left(\bea{cc}\mu&\nu\\
\sigma&\omega\eea\right),E=\left(\bea{c}\kappa\\\omega\eea\right)\eea$&$\bea{l}
R=\left(\bea{cc}\mu&\nu\\\nu&\kappa\eea\right)\\
E=\sigma\eea$\\\hline
2,1,1&$\bea{l}R=\left(\bea{cc}\mu&\nu\\\sigma&\alpha\\\omega&0\eea\right)\\
E=\omega-\alpha\eea$&$\bea{l}R=\left(\bea{cc}\mu&\nu\\
0&\omega\eea\right),
E=\left(\bea{c}\alpha\\\sigma\eea\right)\eea$&$\bea{l}R=\left(\bea{cc}
\mu&\sigma\\0&\nu\eea\right)\\ E=\kappa\eea$\\\hline
2,0,0&$\bea{l}R=\left(\bea{cc}\mu&\nu\\\sigma&\alpha\\\alpha&0\eea\right)\\
E
\texttt{ not existing}\eea$&$\bea{l}R=\left(\bea{cc}\mu&\nu\\
\omega&0\eea\right)\\ E\ \texttt{not
existing}\eea$&$\bea{l}R=\left(\bea{cc}\mu&\nu\\\sigma&0\eea\right)\\
E\ \texttt{not}\\ \texttt{existing}\eea$\\\hline 1,2,1&$\bea{l}
R=\left(\bea{c}\mu\\\nu\\\alpha\eea\right)\\E= (\omega\
\alpha)\eea$&$\bea{l}R=\left(\bea{c}\kappa\\\sigma\eea\right),
E=\left(\bea{cc}\mu&\nu\\
\omega&0\eea\right)\eea$&$\bea{l}R=\left(\bea{c}\mu\\\nu\eea\right)\\
E=(\sigma \ \ 0)\eea$\\\hline
1,1,0&$\bea{l}R=\left(\bea{c}\mu\\\nu\\\alpha\eea\right),  E=
\alpha\eea$&$\bea{l}R=\left(\bea{c}\kappa\\\sigma\eea\right),
E=\left(\bea{c}\mu\\
0\eea\right)\eea$&$\bea{l}  R=\left(\bea{c}\mu\\\nu\eea\right)\\
E=\sigma\eea$\\\hline
1,1,1&$\bea{l}R=\left(\bea{c}0\\\nu\\\alpha\eea\right),\  E=
\omega\eea$&$\bea{l}R=\left(\bea{c}\kappa\\\sigma\eea\right),
E=\left(\bea{c}\mu\\
\nu\eea\right)\eea$&$\bea{l}R=\left(\bea{c}\mu\\\nu\eea\right)\\
E=0\eea$\\\hline
1,0,0&$\bea{l}R=\left(\bea{c}\mu\\\alpha\\0\eea\right)\\  E
\texttt{ not existing}\eea$&$\bea{l}R=\left(\bea{c}\kappa\\\sigma\eea\right)\\
E\ \texttt{not existing}\eea$&$\bea{l}R=\left(\bea{c}\kappa\\\sigma\eea\right)\\
E\ \texttt{not}\\ \texttt{existing}\eea$\\\hline 0,1,0&$\bea{l}R\
\texttt{not existing},\\  E=\alpha\eea$&$\bea{l}R\ \texttt{ not}\\
\texttt{existing},\eea E=\left(\bea{c}\kappa\\\sigma\eea\right)
$&$\bea{l}E=\alpha,\  R\ \texttt{not}\\\texttt{existing}
\eea$\\\hline

\end{tabular}

\newpage

\begin{center}

Table 4. Submatrices $R$ and $E$ of matrices $\beta_4$
\end{center}

 \begin{tabular}{|c|c|c|c|c|}
\hline
& \multicolumn{4}{|c|}{$m,n,\lambda$}\\
\cline{2-5} \raisebox{1.8ex}[0pt][0pt]{$m',n',\lambda'$}
&1,1,0&1,1,1&1,0,0&0,1,0
\\
\hline 1,1,0&$\bea{l}R=\mu\\ E=\nu\eea$&$\bea{l}R=\mu\\
E=\nu\eea$&$\bea{l}
 R=\mu, E\ \texttt{not} \\\texttt{existing} \\\eea$&$\bea{l}E=\mu,
 R\ \texttt{not} \\\texttt{existing} \eea$\\\hline 1,1,1&$\bea{l}R=\mu,
 \\ E=\nu\eea$&$\bea{l}R=\mu\\
E=0 \eea$&$\bea{l}R=\mu,
 E\ \texttt{not} \\\texttt{
existing} \eea$&$\bea{l}E=\mu,
 R\ \texttt{not} \\\texttt{
existing} \eea$\\\hline 1,0,0&$\bea{l}R=\mu,
 E\ \texttt{not} \\\texttt{
existing} \eea$&$\bea{l}R=\mu,
 E\ \texttt{not} \\\texttt{existing} \eea$&$\bea{l}R=\mu, E\texttt{ not}
 \\\texttt{existing}
\eea$&$\bea{l}R \texttt{ and }
 E\\ \texttt{not
existing} \eea$\\\hline 0,1,0&$\bea{l}E=\mu,
 R\ \texttt{not} \\\texttt{
existing} \eea$&$\bea{l}E=\mu,
 R\ \texttt{not} \\\texttt{
existing} \eea$&$\bea{l}R \ \texttt{and }E\\\texttt{not
existing}\eea$&$\bea{l}E=\mu,R \ \texttt{not} \\ \texttt{existing}
\eea$\\\hline
\end{tabular}

\end{document}